\documentclass[10pt,conference]{IEEEtran}

\usepackage{mathptmx}
\usepackage{mathtools}
\usepackage{graphicx}
\usepackage{algorithm}
\usepackage{algpseudocode}
\usepackage{amssymb}
\usepackage{amsmath}
\usepackage{color}	
\usepackage{array}
\usepackage{tabularx}
\usepackage{float}
\usepackage{multirow}
\usepackage{subfigure}
\usepackage{balance}
\usepackage{epsfig}
\usepackage{epstopdf}

\usepackage[strings]{underscore}
\usepackage[labelsep=period]{caption}

\newtheorem{definition}{Definition}       

\newcommand\blfootnote[1]{%
  \begingroup
  \renewcommand\thefootnote{}\footnote{#1}%
  \addtocounter{footnote}{-1}%
  \endgroup
}

\begin{document}
\title{A Comparative Analysis of Materialized Views Selection and Concurrency Control Mechanisms in NoSQL Databases}

\author{\IEEEauthorblockN{Ashish Tapdiya}
\IEEEauthorblockA{Vanderbilt University\\
Email: ashish.tapdiya@vanderbilt.edu}
\and
\hspace*{.1in}
\IEEEauthorblockN{Yuan Xue}
\IEEEauthorblockA{Vanderbilt University\\
Email: yuan.xue@vanderbilt.edu}
\and
\IEEEauthorblockN{Daniel Fabbri}
\IEEEauthorblockA{Vanderbilt University\\
Email: daniel.fabbri@vanderbilt.edu}}

\maketitle

\thispagestyle{plain}
\pagestyle{plain}

\begin{abstract}
Increasing resource demands require relational databases to scale. While relational databases are well suited for vertical scaling, specialized hardware can be expensive. Conversely, emerging NewSQL and NoSQL data stores are designed to scale horizontally. NewSQL databases provide ACID transaction support; however, joins are limited to the partition keys, resulting in restricted query expressiveness. On the other hand, NoSQL databases are designed to scale out linearly on commodity hardware; however, they are limited by slow join performance. Hence, we consider if the NoSQL join performance can be improved while ensuring ACID semantics and \textbf{without} drastically sacrificing write performance, disk utilization and query expressiveness. 

This paper presents the Synergy system that leverages schema and workload driven mechanism to identify materialized views and a specialized concurrency control system on top of a NoSQL database to enable scalable data management with familiar relational conventions. Synergy trades slight write performance degradation and increased disk utilization for faster join performance (compared to standard NoSQL databases) and improved query expressiveness (compared to NewSQL databases). Experimental results using the TPC-W benchmark show that, for a database populated with 1M customers, the Synergy system exhibits a maximum performance improvement of 80.5\% as compared to other evaluated systems.
\end{abstract}
\begin{IEEEkeywords}
Transaction processing, materialized views, NoSQL databases, performance evaluation
\end{IEEEkeywords}
\section{Introduction} 
\label{model}
\blfootnote{\hspace{-8 pt}A short version of this paper appeared in IEEE Cluster Conference 2017. \\ Available Online: http://ieeexplore.ieee.org/document/8048951/}
Application development with relational databases as the storage backend is prevalent, because relational databases provide: a formal framework for schema design, a structured query language (i.e., SQL) and ACID transactions. As applications gain popularity and the database system reaches its resource limits, the architecture must scale up to ensure end-to-end response time (RT). Relational databases are well suited for vertical scaling; however, vertical scaling has known limitations and requires expensive hardware. On the contrary, the new brand of NewSQL and NoSQL databases are known for their ability to scale out linearly \cite{kossmann, ycsb, hstore}. Hence, as the data size and the resource demands increase, application designers can consider transitioning from their relational database to a NewSQL/NoSQL database. Recently Facebook \cite{Fbmsg} and Netflix \cite{Netflix} transformed part of their relational databases to HBase \cite{HBase}.

NewSQL architectures enable a database to scale out linearly while providing ACID transaction guarantees. However, their schema design requires careful consideration when choosing partition keys, since joins are restricted to partition keys only \cite{voltdb}, resulting in limited query expressiveness. Similarly, NoSQL databases can also scale out linearly, but are limited by slow join performance due to the distribution of data across the cluster and data transfer latency, which has also been identified in previous work \cite{asyncView}. Thus, while NewSQL and NoSQL systems allow data stores to scale, their designs sacrifice query expressiveness and join performance, respectively. More generally, there exists a design space that makes trade-offs between performance, ACID guarantees, query expressiveness and disk utilization. 

This paper considers if NoSQL join performance \textbf{can} be improved while ensuring ACID semantics and \textbf{without} drastically sacrificing write performance, disk utilization and query expressiveness. One option for improving the performance of NoSQL workloads is materialized views (MVs), which pre-compute expensive joins \cite{Larson, Yang}. However, deploying MVs on top of a NoSQL store does not guarantee consistency as atomic key based operations allow for the MV's data to be stale relative to the base table \cite{HBase, Bigtable, Accumulo}. Hence, additional concurrency controls such as locking or multi-versioning are required to ensure data consistency.

Standard concurrency control methods, such as locking or multi-versioning, can provide ACID semantics for NoSQL stores with materialized views, but induce performance degradation (i.e., by grabbing many locks or checking multiple versions, respectively) because the concurrency control mechanism and MVs selection mechanism are not designed in tandem. Instead, this paper considers a synergistic design space in which the concurrency control mechanism and MV's selection mechanism operate together such that only a single lock is grabbed per transaction. The proposed system relies on the hierarchical structure of relational data and the workload to inform the views selection mechanism, which can then be leveraged to grab a single lock for MVs and base tables. 
\begin{figure*}[t]
	\centering
		\includegraphics[trim = 20mm 80mm 48mm 15mm, clip, width=.6\textwidth]{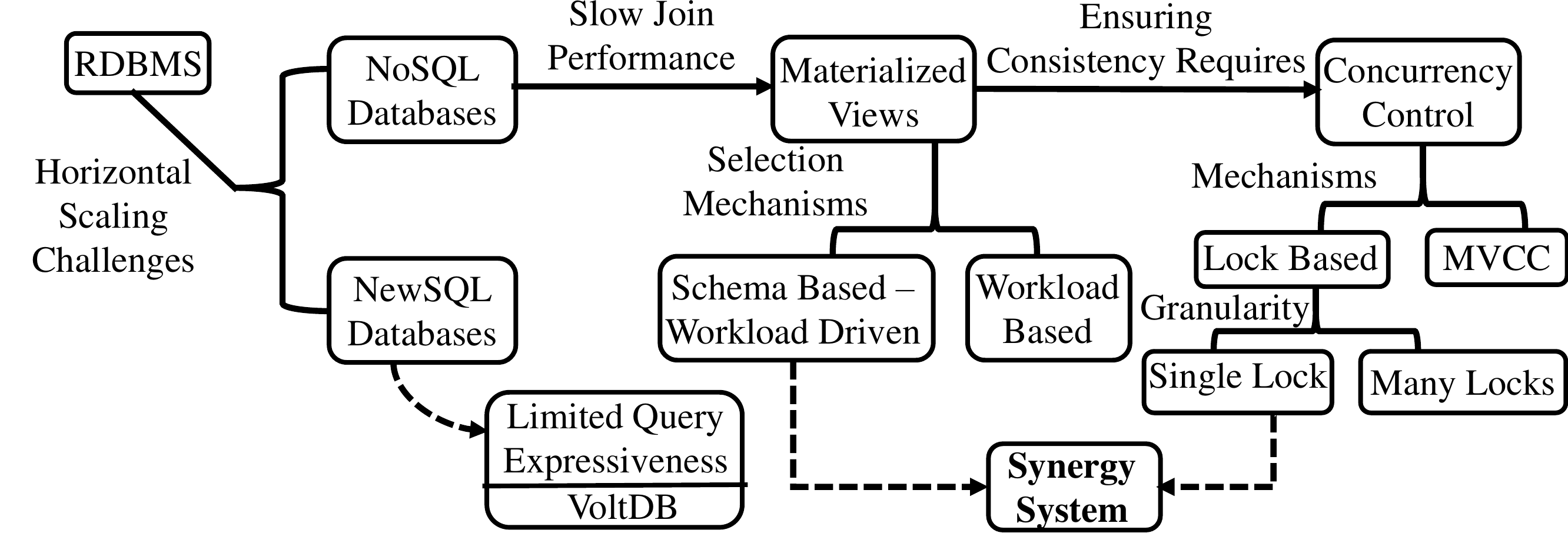}
	\caption{Design choices and decisions in the Synergy System.}
	\label{fig:position}
\end{figure*}
	
In this work, we present the Synergy system that leverages MVs and a light-weight concurrency control on top of a NoSQL database to provide for scalable data management with familiar relational conventions and more robust query expressiveness. Figure \ref{fig:position} presents the design decisions for MVs selection and concurrency control mechanisms in the Synergy system. Synergy harnesses databases' hierarchical schemas to generate candidate MVs, and then uses a workload driven selection mechanism to select views for materialization. To provide ACID semantics in the presence of views, the system implements concurrency controls on top of the NoSQL database using a hierarchical locking mechanism that only requires a single lock to be held per transaction. The Synergy system provides ACID semantics with the read-committed transaction isolation level. Our contributions in this work are as follows: 

\begin{itemize}
	\item We present the design of Synergy system that trades slight write performance degradation and increased disk utilization for faster join performance (compared to standard NoSQL databases) and improved query expressiveness (compared to NewSQL databases).
	\item We propose a novel schema based--workload driven materialized views selection mechanism. 
	\item We implement and evaluate the proposed system on an Amazon EC2 cluster using the TPC-W benchmark.
	\item We compare and contrast the performance of Synergy system with four complementary systems.  
\end{itemize}

\begin{figure}[t]
	\centering
		\includegraphics[trim = 5mm 58mm 45mm 5mm, clip, width=.45\textwidth]{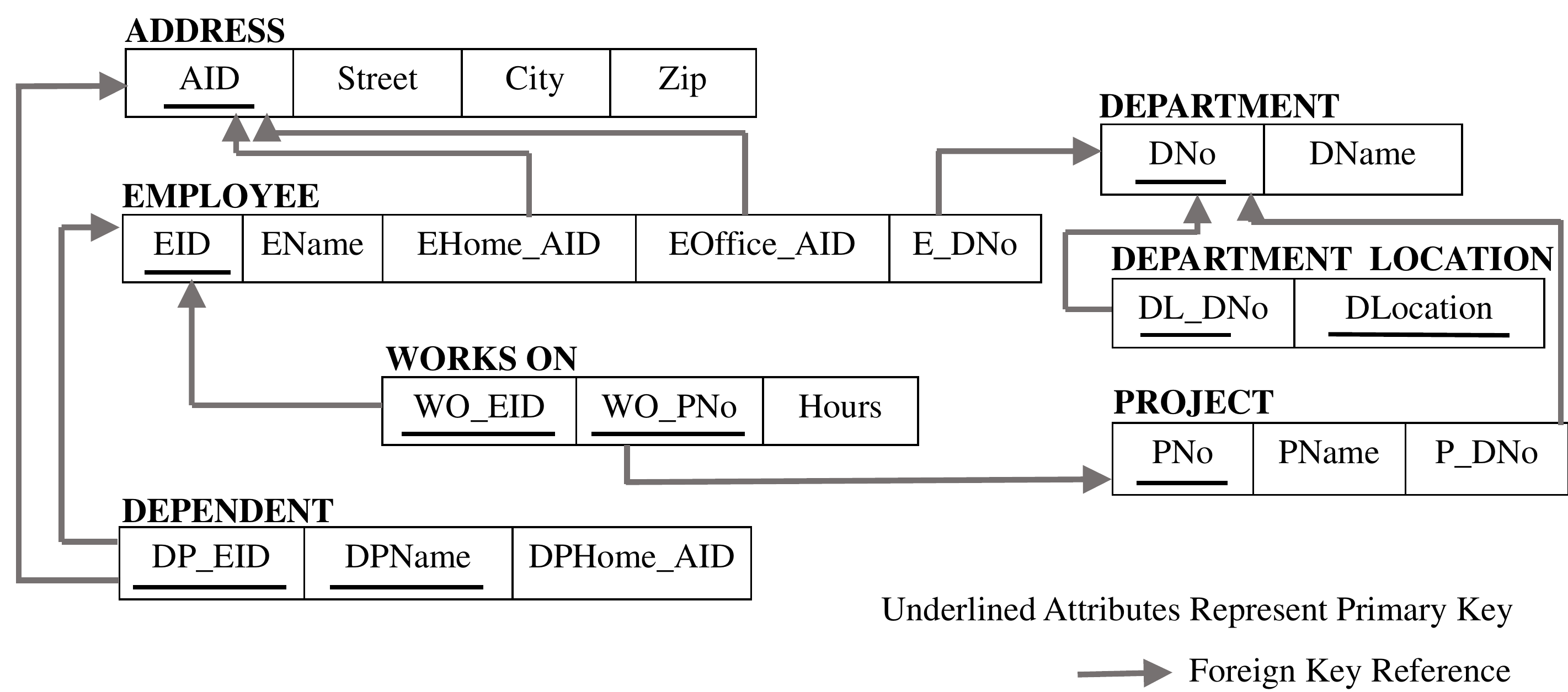}
	\caption{Relations in the Company schema.}
	\label{fig:companyS}
\end{figure}
\section{Background}
\label{sec:backg}
We first review the concepts of a relation, index and schema which are common to both SQL and NoSQL data models. Then, we present a model for the database workload. Finally, we provide an overview of the data store used and its associated SQL skin.
\subsection{Relation, Index and Schema Models}
\label{subsec:model}
\textbf{Relation--} A relation R is modeled as a set of attributes. The primary key of R denoted as PK(R), is a tuple of attributes that uniquely identify each record in R. The foreign key of R denoted as FK(R), is a set of attributes that reference another relation T. A relation can have multiple foreign keys, hence let F(R) denotes the set of foreign keys of R.

\textbf{Index--} In this work we utilize covered indexes that store the required data in the index itself. An index X on a relation R denoted as X(R), is modeled as a set of attributes (s.t. X(R) $\subset$ R). Let X$^{tuple}$(R) denotes a tuple of attributes that the index is indexed upon (s.t. X$^{tuple}$(R) $\subset$ X(R)). The key of an index is a union of attributes in tuples X$^{tuple}$(R) and PK(R), in that order. Since a relation can have multiple indexes, let I(R) denotes the set of indexes on R.

\textbf{Schema--} Using the previous definitions of a relation and an index, a database schema S is modeled as a set of relations and the corresponding index sets, S = \{R$_{1}$, I(R$_{1}$), R$_{2}$, I(R$_{2}$),..., R$_{n}$, I(R$_{n})$\}, where $n$ represents the number of relations in the schema. We use an example Company database for the purpose of exposition. Figure \ref{fig:companyS} depicts the relations in the Company database schema.
\subsection{Modeling Workload}
\label{sec:workload}
A database workload W = \{w$_1$,..., w$_m$\} is modeled as a set of SQL statements, where $m$ is the number of statements.
\subsection{HBase Overview}
\label{sec:hbase}
We use HBase \cite{HBase} for the purpose of exposition and experimentation in this work. It is a column family-oriented distributed database modeled after Google's Bigtable \cite{Bigtable}. HBase organizes data into tables. A table consists of rows that are sorted alphabetically by the row key. HBase groups columns in a table into column families such that each column family data are stored in its own file. A column is identified by a column qualifier. Also, a column can have multiple versions of a data sorted by the timestamp.

The HBase data manipulation API comprises of five primitive operations: Get, Put, Scan, Delete and Increment. The Get, Put, Delete and Increment operations operate on a single row specified by the row key. HBase provides ACID transaction semantics with read-committed isolation level.
\subsection{Phoenix Overview}
\label{sec:phoenix}
Apache Phoenix \cite{Phoenix} is a SQL skin on top of HBase. The client embedded JDBC driver in Phoenix transforms the SQL query into a series of HBase scans and coordinates the execution of scans to generate a standard JDBC result set. The default transaction semantics in Phoenix with base tables only is same as HBase; however, recent integration with Tephra \cite{tephra} enables multi-statement transactions in Phoenix through MVCC. Note, the MVCC transaction support in Phoenix can be turned on/off by starting/stopping Phoenix-Tephra transaction server. Next, we describe the mechanism to perform a \textbf{baseline transformation} from a relational to a NoSQL database. 

\textbf{Baseline Schema Transformation -- } A relation R becomes a relation R$'$  in NoSQL schema with the same set of attributes as R. The row key of R$'$ is a delimited concatenation of the value of attributes in PK(R). Similarly, an index X(R) on a relation R becomes a relation X(R$'$) in NoSQL schema with the same set of attributes as X(R). The row key of X(R$'$) is a delimited concatenation of the value of attributes in the key of X(R). Note that in NoSQL, both for a relation and an index, we assign all attributes to a single column family.  

\textbf{Baseline Workload Transformation -- } Each read statement from the relational workload is added to the NoSQL workload. Each write statement for a relation R that specifies each key attribute in the WHERE clause is added to the NoSQL workload. 

\section{Challenges and Design Choices}
\label{sec:motivate}
Joins are expensive in a NoSQL database due to the distribution of data items across different cluster nodes. It is well understood that MVs improve join  performance by pre-computing and storing results in the database \cite{Larson, Yang, Goldstein}. This observation is verified with TPC-W micro-benchmark  which shows that scanning a MV is significantly faster than the join performance (see Section \ref{sec:microb} for experiment details). Thus, we consider how to incorporate MVs into a NoSQL store, while ensuring consistency.

\subsubsection{Implication of Materialized Views}
NoSQL databases are generally limited to key-based single row operations \cite{HBase, Bigtable, Accumulo}. Hence, to ensure the ACID semantics in the presence of MVs, view maintenance and concurrency controls are required to ensure consistency between the MVs and base tables. The design choices for concurrency control mechanisms include multi-versioning, locking and timestamp ordering. While multi-versioning may seem like a nature fit given HBase and other NoSQL system's temporal key component (i.e., cell values are composed of a row-key, column family, column and time stamp) \cite{Bigtable, Accumulo}, experimental results show that getting and checking additional rows' timestamps decreases performance. Therefore, this result motivates a lock-based concurrency control mechanism to attain the read committed isolation level.

\subsubsection{Lock Number and Granularity}
\label{sec:lockg}
Row level locks and database locks represent the two ends of the locking mechanism spectrum. Database locks degrade system throughput since only a single transaction can access the database at a time. Similarly, acquiring row level locks on individual base tables can be expensive in the presence of MVs in a NoSQL database, since the system may need to acquire a large number of locks for complex queries. Experimental results show that for a modest number of 100 locks, the time to acquire and release locks is 1.3x the response time of the most expensive write transaction in the proposed system (see Section \ref{sec:lock-overhead} and Section \ref{sec:write-overhead}). This observation motivates minimizing the number of locks required per transaction. 

\subsubsection{View Selection Challenges}
The types of MVs that are allowed impact the data store performance in varying ways. Purely workload based MVs selection mechanisms \cite{Agrawal} (schema relationships are oblivious) can result in optimal read performance by allowing for the materialization of a maximum number of joins in the workload (i.e., views constructed with many-to-many joins or non-foreign key joins). While this approach  is well suited for OLAP workloads, it can degrade write performance and increase disk utilization and transaction management costs for the OLTP workloads, especially in a distributed database. In contrast a schema aware--workload driven MVs selection mechanism limits the type of views allowed, resulting in sub-optimal read performance. However, this approach prevents high storage costs and shifts of the bottleneck from read to the write performance. Given the design goal to hold a single lock per transaction across base tables and MVs, this observation motivates us to not allow views with many-to-many joins or joins that do not have key relationships.
\subsection{Design Decisions}
For the Synergy system, we make the following design decisions based off of our analysis of the TPC-W benchmark, which contains many key/foreign-key equi-joins. First, we develop a concurrency control mechanism that leverages the schema's relational hierarchy, grabs one lock per transaction and provides the read committed isolation level. Second, in cooperation with our concurrency control mechanism, the system only materializes key/foreign-key equi-joins, does not materialize joins across many-many relationships, and each base relation may only be assigned to a single relational hierarchy for materialization (so that a single lock must be acquired per transaction). \textit{We believe the synergistic design decisions between the concurrency control and view selection mechanism provides for a novel architecture and substantially differentiates this work from previous works on materialized view selection.}

\begin{table*}[]
\centering
\caption{Qualitative comparison of NoSQL, NewSQL and Synergy systems.}
\label{tab:compare-systems}
\scalebox{0.7}{
\begin{tabular}{|c|c|c|c|c|}
\hline
                                                                     & \textbf{Scalability} & \textbf{Query  Expressiveness}                                                                          & \textbf{Transaction Support}                                                              & \textbf{Disk Utilization}                                                        \\ \hline
\textbf{\begin{tabular}[c]{@{}c@{}}NoSQL\\   (HBase)\end{tabular}}   & Linear scale out     & SQL                                                                                         & ACID with Snapshot Transaction  Isolation                                                                & Higher than NewSQL                                                                           \\ \hline
\textbf{\begin{tabular}[c]{@{}c@{}}NewSQL\\   (VoltDB)\end{tabular}} & Linear scale out     & \begin{tabular}[c]{@{}c@{}}SQL with joins limited to \\ partition keys\end{tabular}                     & ACID with Serializable Transaction Isolation                                                                                      & Lowest \\ \hline
\textbf{Synergy}                                                & Linear scale out     & \begin{tabular}[c]{@{}c@{}}SQL with MVs limited to \\ Key/Foreign-Key joins\end{tabular} & \begin{tabular}[c]{@{}c@{}}ACID with Read Committed \\ Transaction Isolation\end{tabular} & Highest                                                                          \\ \hline
\end{tabular}
}
\end{table*}

\section{System Overview}
\label{sec:overview}
\begin{figure}[t]
	\centering
		\includegraphics[trim = 10mm 67mm 10mm 15mm, clip, width=.5\textwidth]{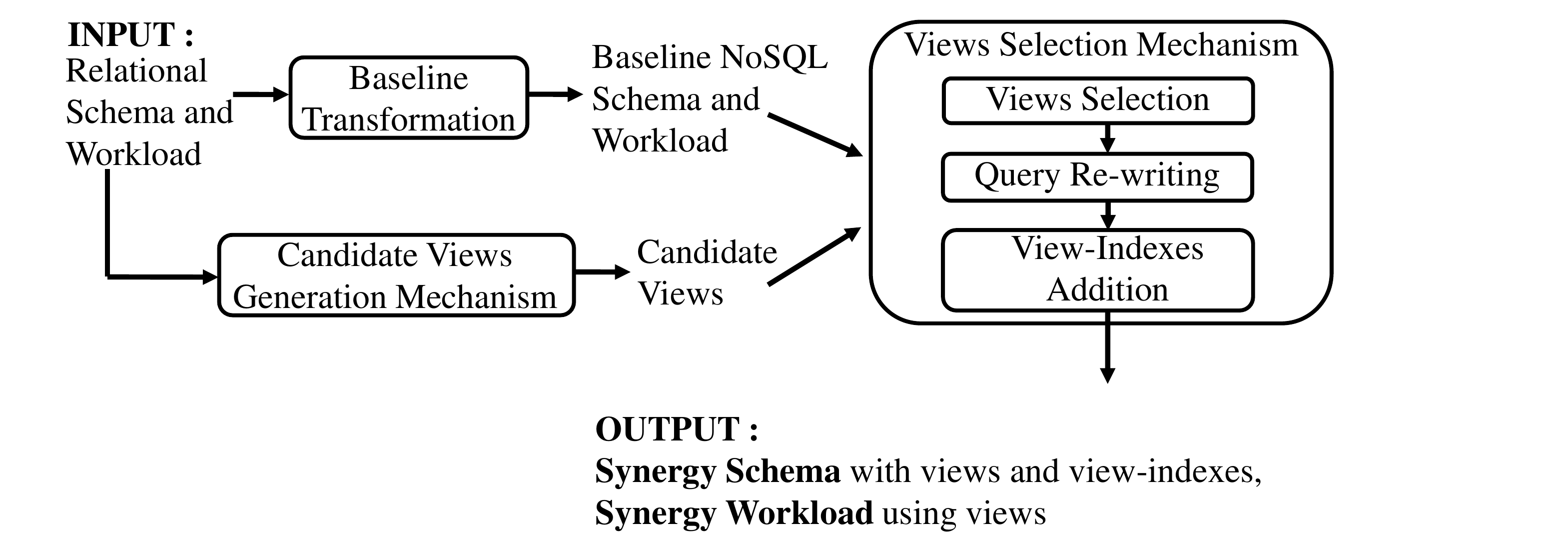}
	\caption{Synergy system overview.}
	\label{fig:workflow}
\end{figure}

In this section we provide an overview of the \textbf{Synergy system}, as depicted in Figure \ref{fig:workflow}. The objective of our system is to design a scalable and high performance NoSQL database while ensuring  the ACID semantics.

We first perform a baseline transformation of the input relational database to a NoSQL database using the mechanism described in Section \ref{sec:phoenix}. Due to the slow join performance in the baseline transformed database system, we decide to use MVs. We use the \textbf{candidate views generation mechanism} to create a list of potential views to materialize based on the database's hierarchical structure. Next, we use a \textbf{workload driven view selection mechanism} to select views from the candidate set. Then, we re-write the workload using selected views as needed. To ensure high read performance, we supplement the schema with additional view-indexes. To ensure ACID semantics in the presence of views, we implement a concurrency control layer on top of HBase (as described in Section \ref{sec:arch}), which is able to grab a single lock per transaction, while providing the read committed transaction isolation level.

\textbf{System Limitations --} The Synergy system only materializes key/foreign-key equi-joins. In addition, Synergy system is restricted to single SQL statement transactions. In agreement with our design decision of single lock per transaction, write statements that do not specify all key attributes and affect multiple base table rows are not supported. The Synergy system does not enforce foreign key constraints. The transaction isolation level in the Synergy system is limited to read committed. In addition, the Synergy system can only be used with NoSQL data stores that trade availability for strong consistency in presence of network partition (CP model from the CAP theorem \cite{cap}).

\section{Generating Candidate Views}
\label{sec:treem}
In this section we present a mechanism to create candidate views for the materialization of equi joins in the workload. We observe that the joins are slow in a NoSQL database (see Section \ref{sec:motivate}). Hence, materializing the joins in the workload as views can improve the query performance. We harness the schema's structure to identify the candidate views, in particular the key/foreign-key relationships. We begin by presenting formal definitions for schema relationships and views.

\begin{figure}[t]
	\centering
		\includegraphics[trim = 4mm 17mm 18mm 8mm, clip, width=.48\textwidth]{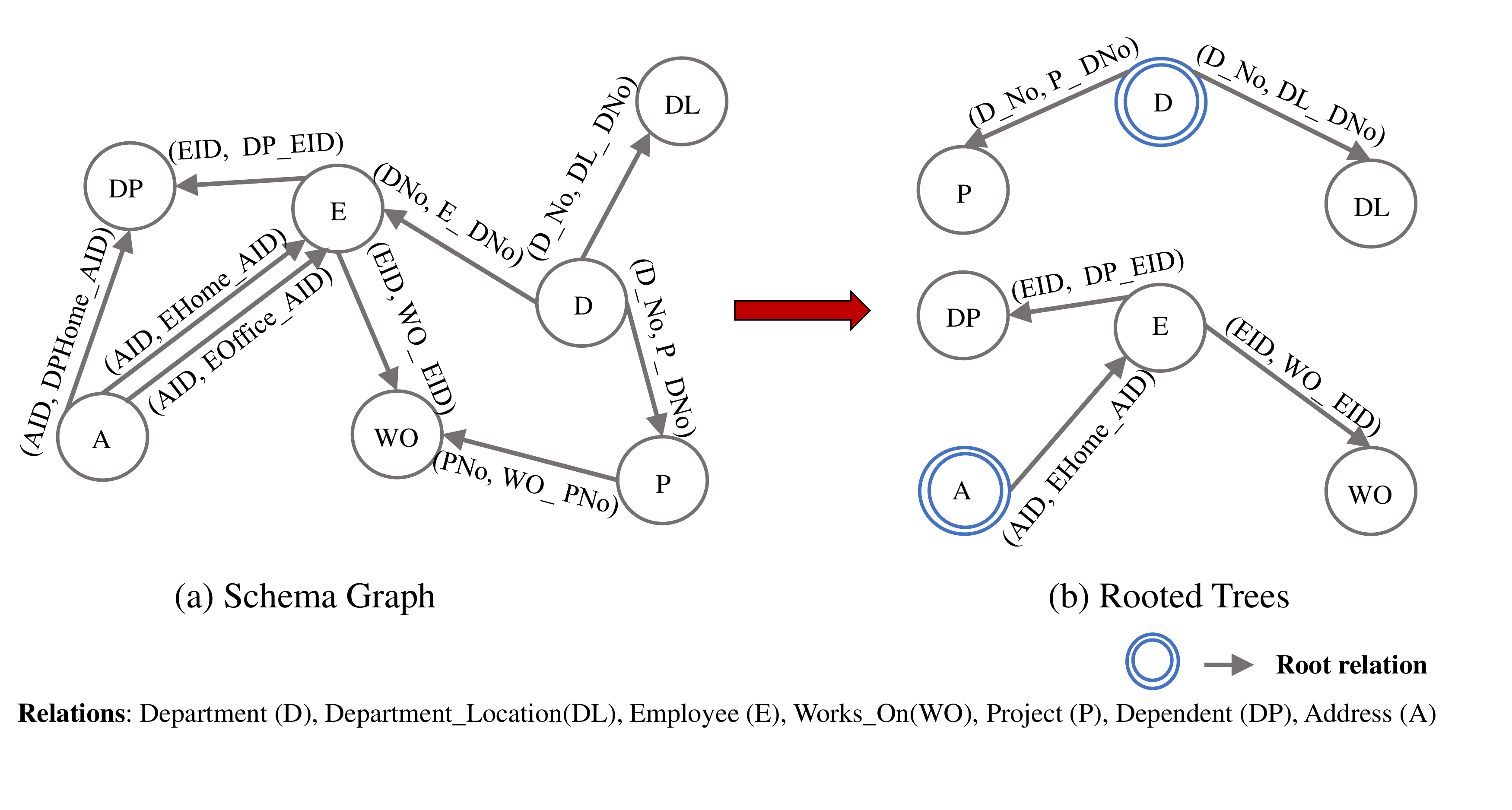}
	\caption{Input and output of the candidate views generation mechanism for the Company database with roots set Q$_{company}$ = $\{\textit{Address, Department}\}$.}
	\label{fig:companyG}
\end{figure}

We assume that the input schema S is normalized and free from both simple and transitive circular references, to limit the scope of this work. We model the relationships in S as a directed graph G=(H,E). The vertices in G represent the relations in S and edges encode the key/foreign-key relationship between relations. An edge exists between relations R$_i$ and R$_j$, if they are related as described by:

\begin{definition}[Schema Relationships]
The relationship between relations R$_i$ and R$_j$, denoted as $R_i \leftarrow R_j$, exists iff FK$_{k}$(R$_i$) references PK(R$_j$), where FK$_{k}$(R$_i$) $\in$ F(R$_i$)
\end{definition} 
Figure \ref{fig:companyG}(a) depicts the schema graph corresponding to the relations in the Company database schema in Figure \ref{fig:companyS}. Next, we define an edge and a path in the schema graph. 
\begin{definition}[Edge in Schema Graph]
A directed edge e$_i$ in a schema graph from a relation R$_i$ to a relation R$_j$ is represented as a \textit{(PK,FK)} tuple where PK is the primary key of R$_i$ and FK is the foreign key of R$_j$.
\end{definition} 
\begin{definition}[Path in Schema Graph]
A path between relations R$_i$ and R$_j$ in a schema graph is modeled as an alternating sequence of relations and directed edges between the relations, [R$_i$,e$_i$,...,e$_{j-1}$,R$_j$]. The alternating sequence begins and ends in a relation. 
\end{definition} 
Database schemas have a hierarchical structure; hence, we can choose a set of relations in the schema graph as roots to create rooted trees. Next, we define a rooted tree.
\begin{definition}[Rooted Tree]
A rooted tree T is a directed graph composed of a subset of nodes and edges from the schema graph in which there exists a root node, and unique paths from the root node to each non-root node.  
\end{definition}
We use rooted trees to identify the candidate views. Next, we define a candidate view.
\begin{definition}[Candidate View]
A candidate view V is a path in a rooted tree. A view is stored physically as a relation. The attributes of V is a set union of attributes of relations in V and the key of V denoted as PK(V) is the key of the last relation in the view. Also, a view-index has the same definition and semantic as a table index.  
\end{definition}
\subsection{Roots Selection}
Each view has a single root. The set of roots Q for a schema S is a subset of relations in S. Q can either be provided by the database designer or it can be learned in an automated manner. In this work, we assume that the database designer provides Q. Note that the automated selection of roots is a separate problem and can be addressed independently. 
\begin{figure*}
\hfill
\subfigure[\small Schema DAG]{\includegraphics[trim = 35mm 45mm 23mm 30mm, clip, width=5cm]{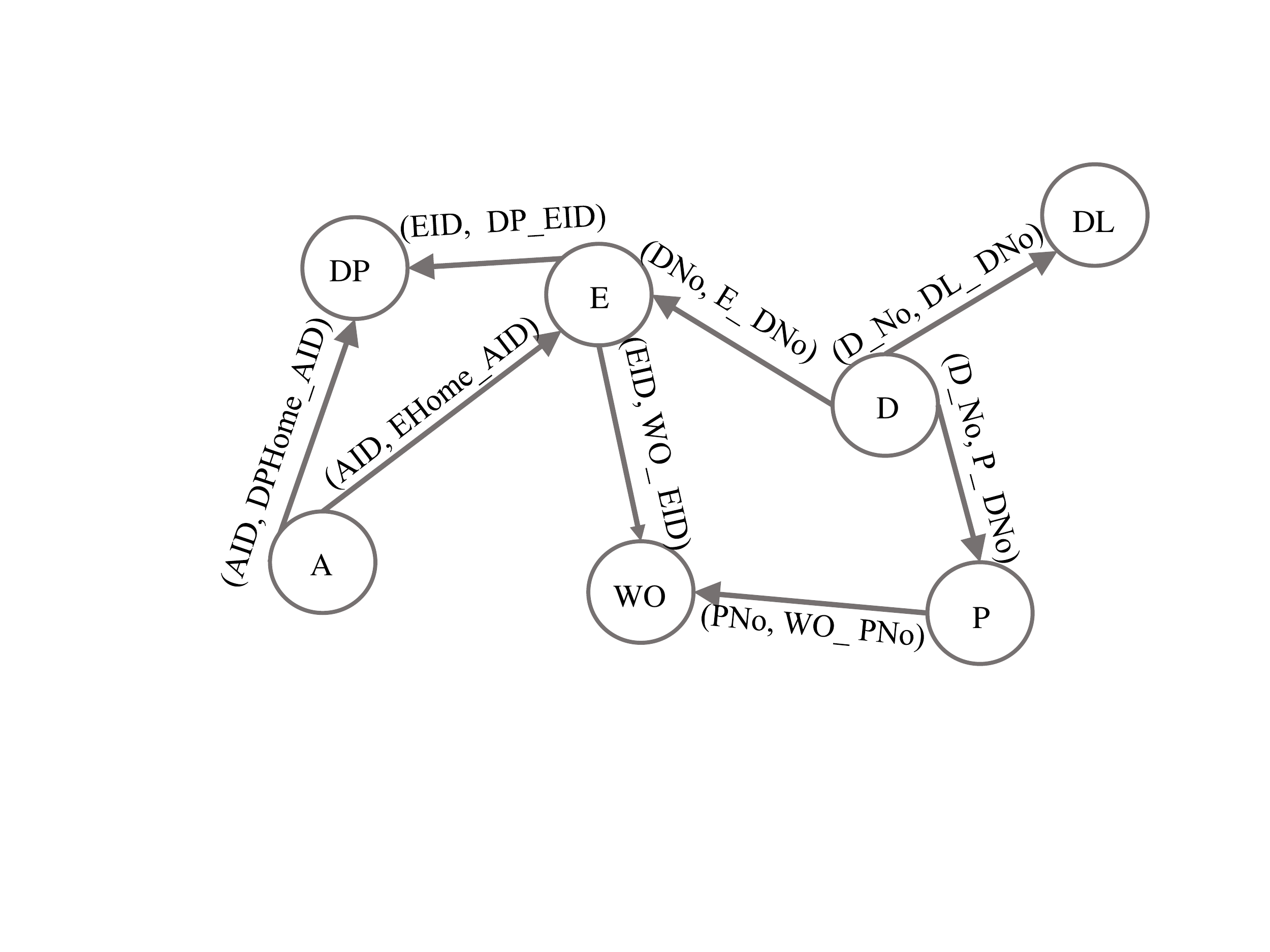} \label{fig:dag}}
\subfigure[\small A topological ordering of schema DAG]{\includegraphics[trim = 5mm 45mm 1mm 10mm, clip, width=6cm]{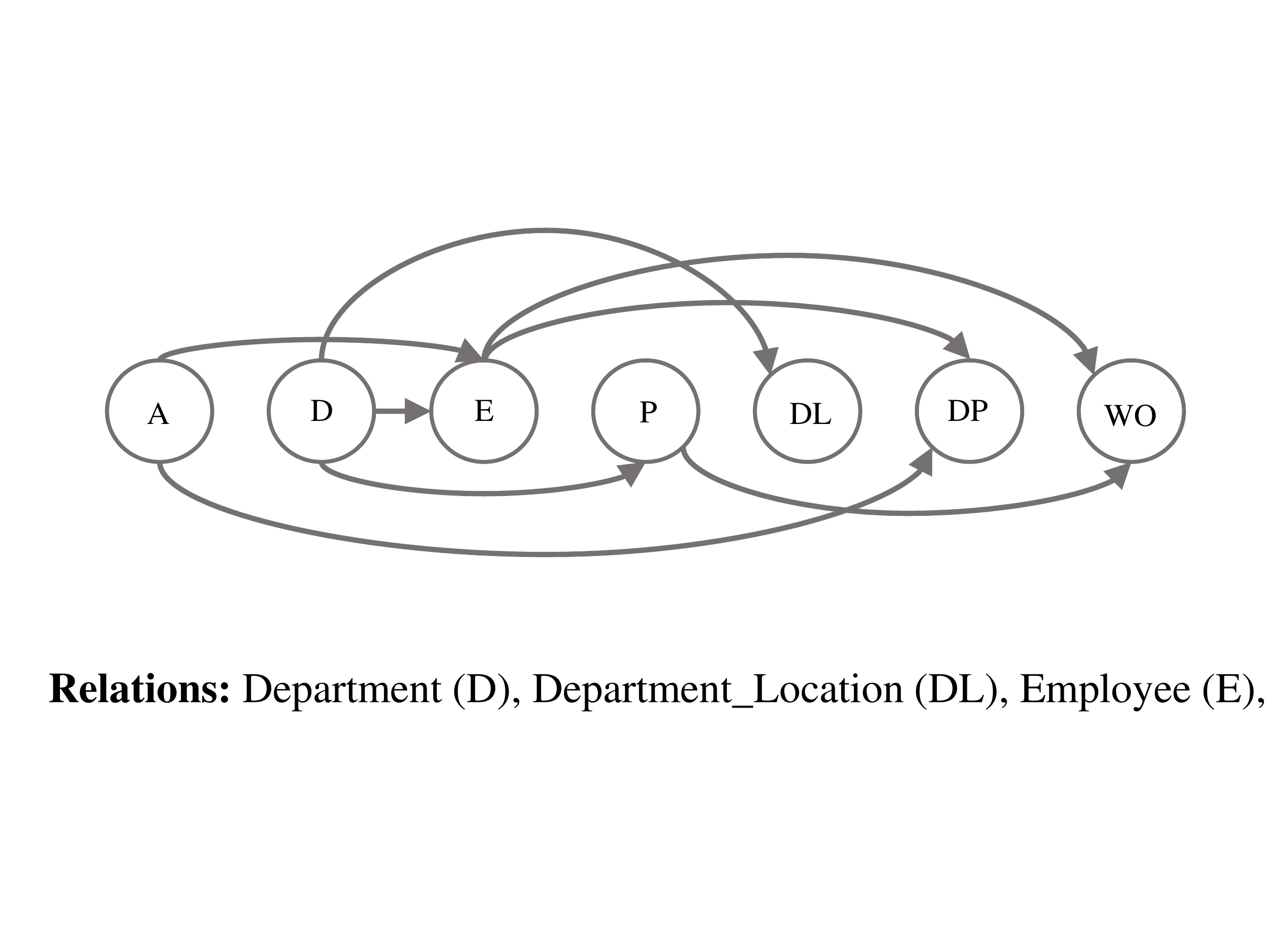} \label{fig:topological}} 
\hspace*{-.6 pc}
\subfigure[\small Rooted Graphs]{\includegraphics[trim = 5mm 30mm 0mm 10mm, clip, width=6cm]{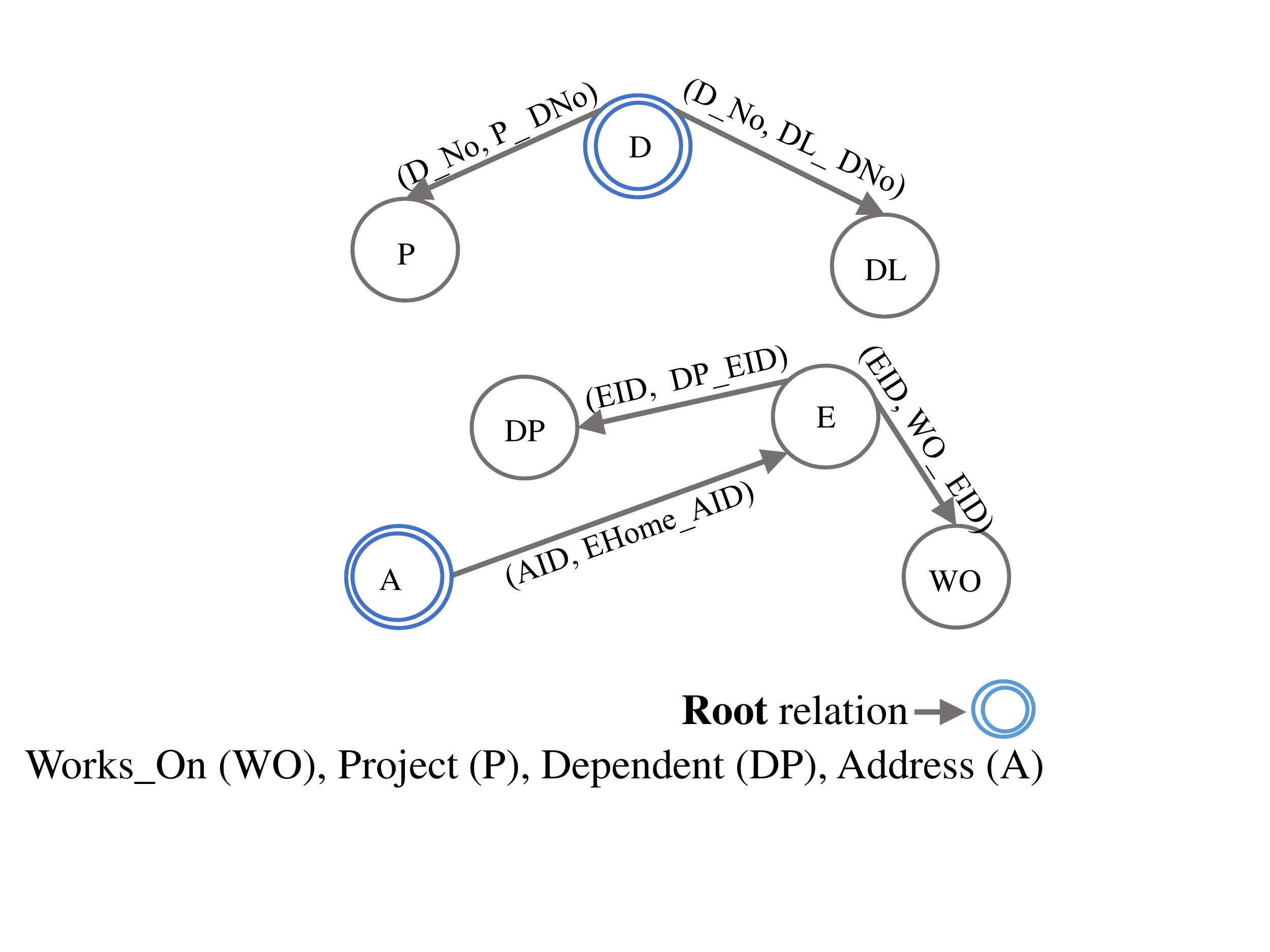} \label{fig:subdags}}
\caption{\small Intermediate results of the candidate views generation mechanism for the Company database with roots set Q$_{company}$ = $\{\textit{Address, Department}\}$.}
\label{fig:inter}
\end{figure*}
\subsection{Candidate Views Generation Mechanism}
The goals of the candidate views generation mechanism are as follows:
\begin{itemize}
	\item \textbf{Assign each non-root relation in the schema graph to at most one root.} This enables us to hold a single lock on the root relation's row key while ensuring ACID semantics.
	\item \textbf{Select a single path between the root and each non-root relation assigned to it.} If there are multiple paths between any pair of relations, then the relationship can be one-many or many-many. However, recall that we define many-many relationship as a join materialization boundary. Therefore, to ensure a one-many relationship, we should select a single path.
\end{itemize}
\subsubsection{Mechanism Overview}
In this section, we present the overview of the mechanism to generate the candidate views. We first \textbf{transform the input schema graph into a directed acyclic graph} (DAG) to ensure at most one direct path between any pair of relations in the graph. Thereafter, we \textbf{identify a topological ordering of the relations} in the schema DAG. Next, we use the topological order to iteratively examine and \textbf{assign each non-root relation to a root} by selecting a path from the root to the non-root relation. Following the assignment of schema relations to roots, a rooted graph is created for each root relation. Finally, we \textbf{transform each rooted graph into a rooted tree} to ensure a single path between the root and each non-root relation. The output of the mechanism is a set of rooted trees and each unique path in a rooted tree represents a candidate view.
\subsubsection{Mechanism Description}
\label{sec:mechanism}
In this section, we describe the candidate views generation mechanism in detail using our continuing example of the Company database. We use Q$_{company}$ = \{\textit{Address,Department}\} as the roots set. In addition, we use a synthetic workload for the purpose of exposition with W$_{Company}$ = \{w$_1$,w$_2$,w$_3$\},
\vspace{.4 pc}

\hspace*{-.1 in}\textbf{W$_1$:} Get address details of an employee\\
\hspace*{.1 in}SELECT * FROM Employee as e, Address as a\\
\hspace*{.1 in}WHERE a.AID = e.EHome$\_$AID and e.EID = ?\\

\hspace*{-.1 in}\textbf{W$_2$:} Get all the employees and their hours who work in a department.\\
\hspace*{.1 in}SELECT * \\
\hspace*{.1 in}FROM Department as d, Employee as e, Works_On as wo\\
\hspace*{.1 in}WHERE d.DNo = e.E$\_$DNo and e.EID = wo.WO$\_$EID\\
\hspace*{.1 in}and d.DNo = ?\\

\hspace*{-.1 in}\textbf{W$_3$:} Get all the employees who work a certain number of hours.\\
\hspace*{.1 in}SELECT * FROM Employee as e, Works_On as wo\\
\hspace*{.1 in}WHERE e.EID = wo.WO_EID and wo.Hours = ? 
\vspace{.4 pc}

\textbf{Heuristic--} During the different steps of the mechanism, we use a heuristic based approach to select a candidate from a set. We choose the \textbf{number of overlapping joins} as a simple workload aware heuristic to assign a weight to each candidate. Note that other heuristics can be used seamlessly with the mechanism.
\vspace{.4 pc}

\hspace*{-.12 in}\textbf{Input:} Schema graph G, workload W and the roots set Q.

\hspace*{-.13 in}\textbf{Output:} Set of rooted trees.

\hspace*{-.13 in}\textbf{Steps:}
\begin{enumerate}
	\item \textbf{Transform input graph to DAG}: In the first step we transform the input schema graph G into a DAG. We achieve this by selecting and keeping at most one edge between any pair of nodes in the schema graph. 

We use our heuristic to assign a weight to each candidate edge. Then, we select the edge with maximum weight and remove the rest. For example, we remove the (AID, EOffice_AID) edge from the schema graph in Figure \ref{fig:companyG}(a) to generate the schema DAG depicted in Figure \ref{fig:dag}.
	\item \textbf{Topologically order relations in the DAG}: Next, we identify a linear ordering of the relations in DAG such that for every directed edge from relation R$_i$ to R$_j$, R$_i$ comes before R$_j$ in the ordering. Figure \ref{fig:topological} represents a topological ordering of the schema DAG presented in Figure \ref{fig:dag}. 
	\item \textbf{Assign relations to roots}: Next, in the topological order, we examine each non-root relation in the schema DAG and decide upon its assignment to a root by executing the following steps:
	\label{step:3}
			\begin{enumerate}
			\item \textbf{Identify paths:} We identify paths in the DAG from each root relation to the non-root relation.
			\item \textbf{Select a path:} Next, we utilize our heuristic to assign a weight to each path. Then, we iterate over the paths in the sorted order by weight until we find a path that includes a single root relation and none of the relations on the path are assigned to a root other than the root present in the path.
			\item \textbf{Add path:} Then, we add the selected path to the rooted graph created for the root in the path.
			\end{enumerate}
Figure \ref{fig:subdags} depicts the rooted graphs generated for the Company database.
	\item \textbf{Transform rooted graphs to rooted trees}: Next, we transform the rooted graphs created in step \ref{step:3} into rooted trees. We first identify a topological ordering of the non-root relations in the rooted graph. We repeat the next step while we have relations left in the topological ordering.
	\label{step:4}
\begin{enumerate}
	\item \textbf{Select a Path}: Using the rooted graph we identify paths between the root relation and the last relation in the topological ordering. Next, we assign a weight to each path using our heuristic. Then, we select the path with maximum weight and add it to the rooted tree. Thereafter, we remove all non-root relations in the path from the topological ordering and continue.
\end{enumerate}
\end{enumerate}
\begin{figure*}[t]\centering
\hspace*{-2.4 in}
\begin{minipage}[t]{0.24\textwidth}\centering
\subfigure { 
\includegraphics[trim = -10mm 35mm 0mm 10mm, scale=.28]{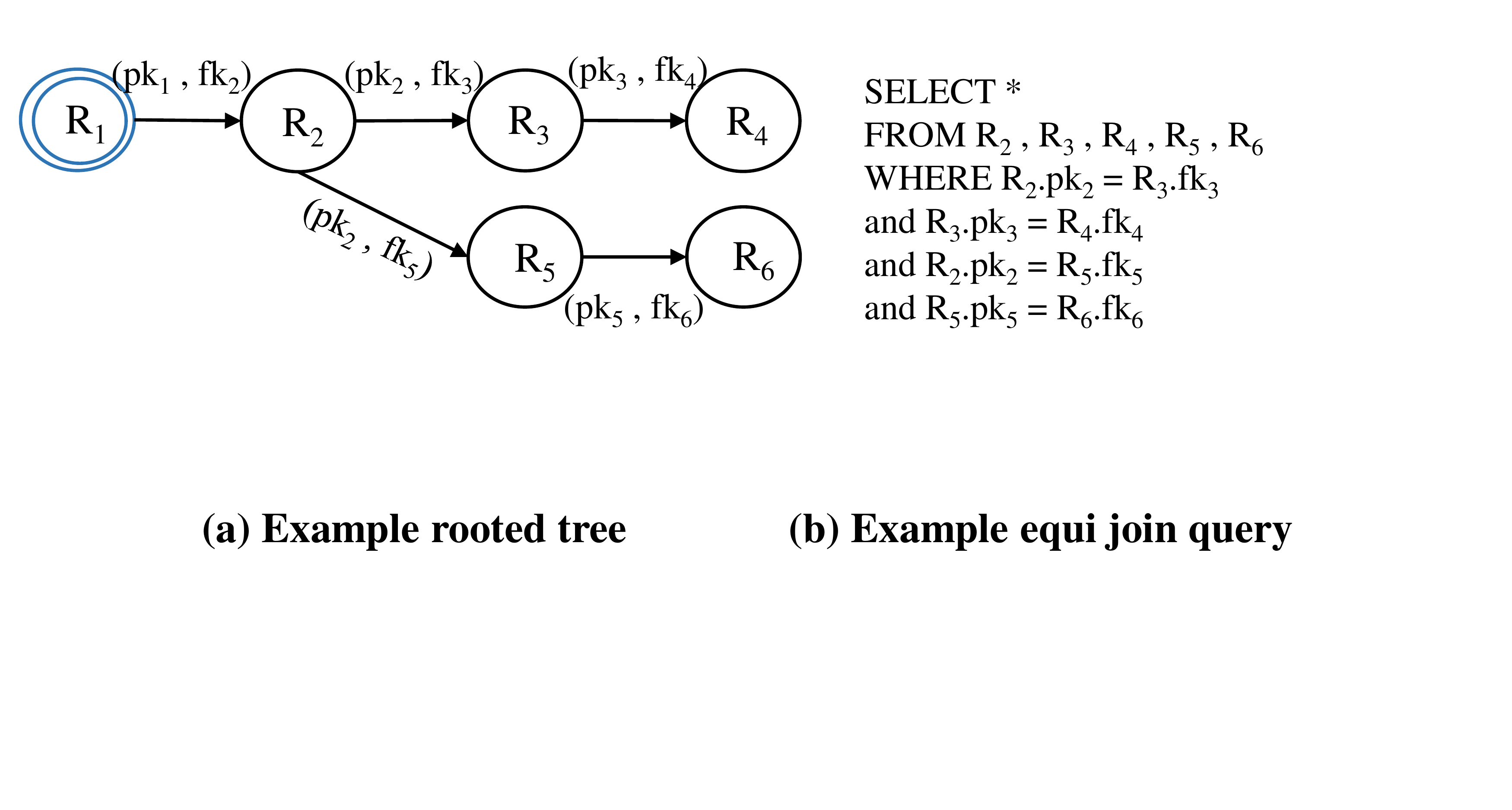}
\label{fig:processdomu}
}
\end{minipage}
\hspace*{1.7 in}
\begin{minipage}[t]{0.24\textwidth}\centering
\subfigure { 
\includegraphics[trim = -8mm 35mm 0mm 10mm, scale=0.28]{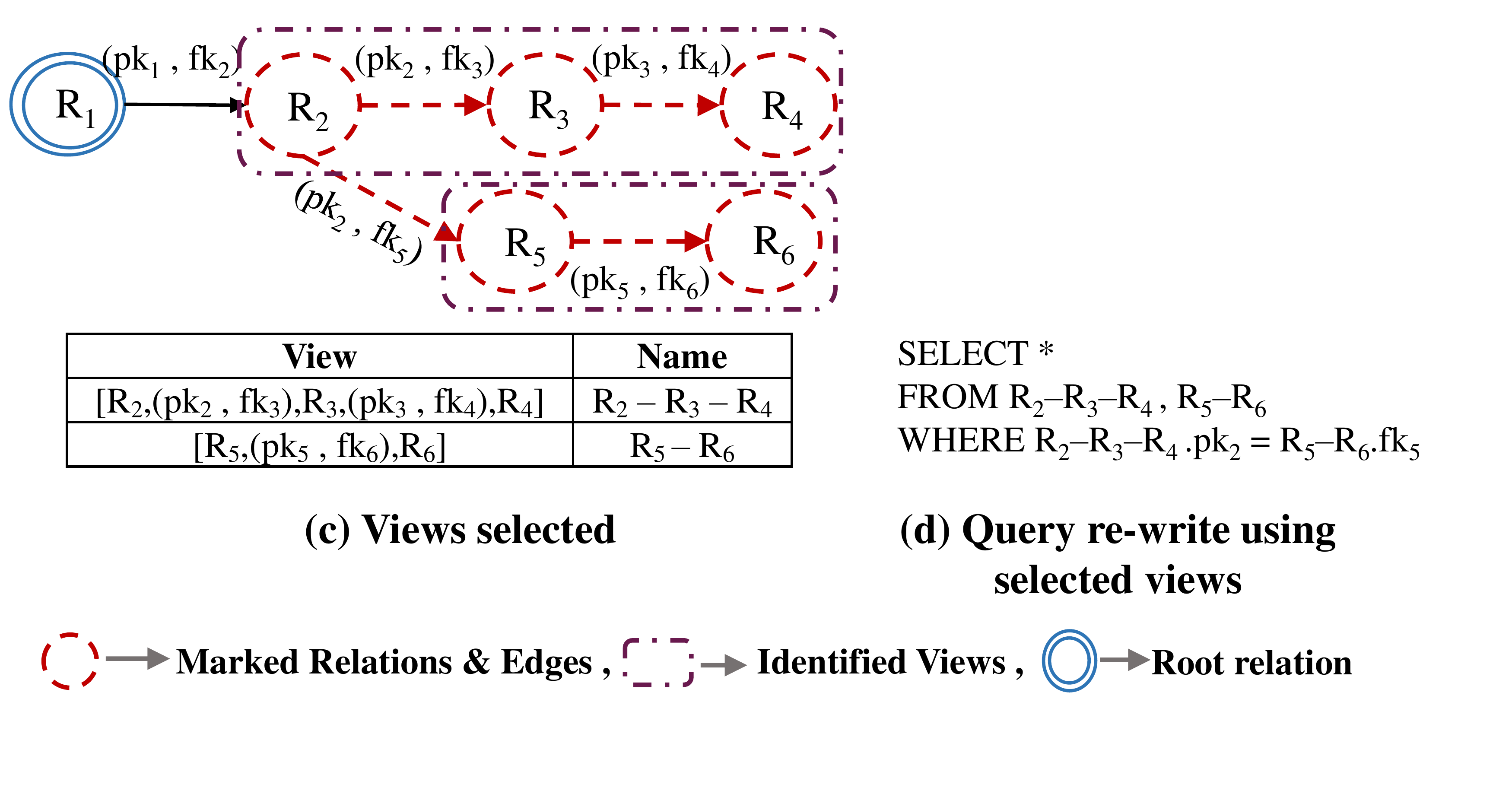}
\label{fig:processdom0}
}
\end{minipage}
\caption{\small Illustration of view selection and query re-writing procedure for an example equi join query using an example rooted tree.}
\label{fig:exViewCreation}
\end{figure*}
Note that in step \ref{step:3}, we examine non-root relations of a schema DAG in the forward topological order to give each non-root relation a chance to be assigned to any root that has a path to it. Conversely, in step \ref{step:4}, we examine non-root relations of a rooted graph in the reverse topological order to keep the paths that will allow materialization of maximum number of joins in the workload. Following the candidate views generation mechanism, a rooted tree is generated for each root in Q. Figure \ref{fig:companyG}(b) depicts the set of rooted trees generated for the Company database.
\subsubsection{Discussion}
The proposed candidate views generation mechanism is a heuristic based approach; hence, does not guarantee materialization of optimal number of joins in the workload. In addition, the usability of generated candidate views for join materialization is dependent on roots selection.

\section{Views Selection Mechanism}
\label{sec:selection}
In this section we describe our procedures for views selection from the candidate set and re-writing queries using selected views. Similar to \cite{Agrawal}, we use a workload driven views selection mechanism. We also illustrate our method for supplementing the schema with additional indexes to ensure query performance.
\subsection{Views Selection}
The high resource requirement and the expensive nature of joins in a NoSQL database (see Section \ref{sec:motivate}) provides us with the motivation to materialize as many joins in the workload as possible to ensure low request response times and high system throughput. We use a workload driven approach to select views. We iteratively examine each equi join query in the workload and select views for it. Next, we describe our procedure to select views for a given query.

\textbf{Views selection for a Query--}We harness the rooted trees and the query syntax to select views for a query. To illustrate the procedure, we use the example rooted tree and the example query depicted in Figures \ref{fig:exViewCreation}(a) and \ref{fig:exViewCreation}(b) respectively. We begin the procedure with un-marked rooted trees. Then, we use the join conditions in the query to mark the relevant edges and participating relations in the rooted trees. Figure \ref{fig:exViewCreation}(c) depicts the marked edges and relations in the example rooted tree. Next, we examine each rooted tree to identify the views to be selected for the query.

For a given rooted tree, we iteratively choose a path until no new path can be chosen. During each iteration, path selection is done using two rules: 1) all the nodes and edges in the path are marked, and 2) the path starts in a marked node that has no incoming marked edge and ends in either a leaf node or a node that has no outgoing marked edge. Then, we select the chosen path as a view. Next, we un-mark the participating relations of the path and outgoing edges of the participating relations, in the rooted tree. Thereafter, we continue with the next iteration. Figure \ref{fig:exViewCreation}(c) depicts the views selected for the example query. 

\textbf{Final View Set--} After processing the entire workload, we add the set of all selected views to the schema. 

\textbf{Limitations--} 1) We select views only for the equi join queries in the workload. 2) Searching the space of all syntactically relevant views is not feasible in practice \cite{Agrawal}; hence, similar to \cite{Agrawal}, our views selection procedure is heuristic based and does not necessarily select the optimal set of views. 3) A views selection procedure that can take advantage of view sharing opportunities across different queries is part of our future work. 4) Currently we do not pass a storage constraint to our views selection algorithm; however, it can be easily adapted to use storage constraint in presence of a cost based query optimizer.
\subsection{Query Re-writing} 
Following views selection, we re-write queries using selected views. We iteratively examine each equi join query in the workload and re-write it using the views selected for it. To re-write a query, we replace the constituent relations of a view with the view. In addition, we remove the join conditions for which both participating relations belong to a single view. Figure \ref{fig:exViewCreation}(d) depicts the example query re-written using selected views. 
\subsection{Additional View Indexes}
Unfortunately, in certain scenarios query execution times can be high despite the use of views. Consider a case in which the query using the view has a filter on an attribute other than the attribute that the view is indexed upon. Then, to prepare the query response, we have to scan the entire view. This can be expensive, depending on the size of the view. Hence, to improve the performance of workload queries that use views, we supplement the schema with additional indexes.

For each view, we examine each conjunctive query that uses this view and decide whether to add a view-index or not. If the query only has filters on one or more view attributes that neither the view nor any of its indexes are indexed upon, then we add a view-index indexed upon a filter attribute to the schema. Note that in this work we do not recommend indexes on base tables and assume that the input schema has necessary base table indexes.

\section{View Maintenance Mechanism}
\label{sec:maintenance}
In this section, we describe the mechanism for view maintenance as the underlying base tables are updated. 
For each type of write statement we present: 1) an applicability test to determine if a base table update applies to a view and 2) a tuple construction procedure to prepare tuples for the view update upon a base table update.
\subsection{Insert Statement}
\subsubsection{Applicability Test}
A base table \textit{insert} for a relation R$_i$ applies to a view V$_i$ iff R$_i$ is the last relation in V$_i$'s sequence of relations.
\subsubsection{Tuple Construction}
Insertion into a view upon a base table insert may require reading tuples from the base tables to construct the view tuple. For a base table insert that applies to a view with $k$ relations, we need to read related tuples from $k-1$ base tables to construct the view tuple. We utilize the key/foreign-key relationships between view relations to sequentially read the base table tuples, starting with relation R$_{k-1}$ and ending in relation R$_1$. Then, we construct the view tuple using previously read tuples and the insert statement. Notice that the time to create a view tuple increases linearly with the number of relations in the view and is independent of the cardinality ratios between the relations.   
\subsection{Delete Statement}
\subsubsection{Applicability Test}
A base table \textit{delete} for a relation R$_i$ applies to a view V$_i$ iff R$_i$ is the last relation in V$_i$. Note that we do not perform cascading deletes.
\subsubsection{Key Construction}
To delete a view tuple upon a base table \textit{delete}, we use the base table key provided with the delete statement. 
However, to delete the view index tuple, we need to first construct the index key to issue a delete upon. Hence, we first read the tuple from the view using the base table key in the delete statement. Then, we use the attributes in the read tuple to construct the index key and issue the delete. Notice that the time to construct a view index key is constant.
\subsection{Update Statement}
\subsubsection{Applicability Test}
A base table \textit{update} for a relation R$_i$ applies to a view V$_i$ iff R$_i$ is in V$_i$'s sequence of relations.
\subsubsection{Tuple Construction}
Unfortunately, updating the view upon a base table update can be expensive if the view is not indexed on the key of the update statement, since we need to either join the base tables or scan the entire view for the tuple construction. To efficiently prepare view updates, we supplement the schema with additional indexes based on the workload. Due to space concerns, we omit the details.

\begin{figure}[t]
	\centering
		\includegraphics[trim = 0mm 15mm 20mm 12mm, clip, width=.45\textwidth]{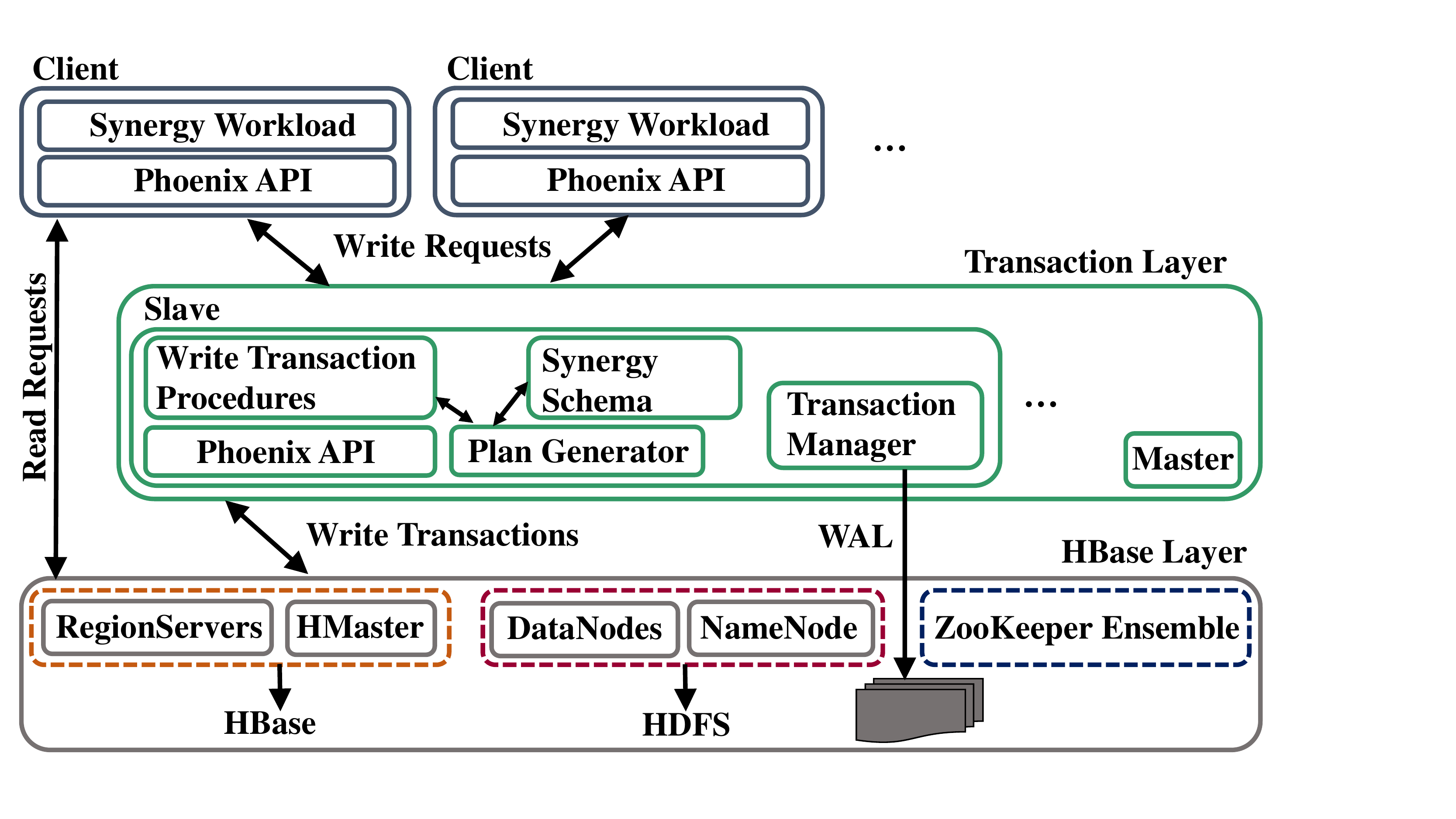}
	\caption{Synergy System Architecture Overview.}
	\label{fig:arch}
\end{figure}
\section{System Architecture}
\label{sec:arch}
In this section we describe the Synergy system architecture. The Synergy system comprises of \textbf{HBase layer}, \textbf{clients} and the \textbf{Transaction layer} as depicted in Figure \ref{fig:arch}. 

\textbf{HBase layer--} The Synergy system harnesses HBase layer as the distributed data storage substrate. The HBase layer comprises of HBase, HDFS and ZooKeeper components. We refer the reader to \cite{HBase} for the role and description of each component shown in Figure \ref{fig:arch}. 

\textbf{Clients--} The clients utilize Phoenix API to execute read and write statements in the workload. A client sends a read request directly to the HBase layer. On the contrary, a write request is sent to the Transaction layer, followed by a synchronous wait for a response.

\textbf{Transaction Layer--} The Synergy system employs the Transaction layer for implementing ACID transaction support on top of the HBase layer. The Transaction layer is a distributed, scalable and fault tolerant layer that comprises of a \textit{Master} node and one or more \textit{Slave} nodes. The \textit{Slave} nodes receive and process \textit{write requests} from clients. Each slave node has a transaction manager that implements a write ahead log (WAL) for recovery and durability. The WAL is stored in HDFS. Upon receiving a request, the transaction manager first assigns a transaction $id$ to the statement and then appends the statement in WAL along with the assigned $id$. Then, a transaction procedure utilizes Phoenix API to execute the transaction. Finally, a response is sent back to the client. The \textit{Master} node is responsible for detecting slave node failures and starting a new slave node to take over and replay the WAL of a failed slave node.

\subsection{Lock Implementation}
\textbf{Logical Locking--} Recall that we restrict the write workload to statements that specify all key attributes (see Section \ref{sec:overview}) and decide to employ hierarchical locking as the concurrency control mechanism (see Section \ref{sec:motivate}). Hence, to update a row for a relation in a rooted tree, we acquire the lock on the key of the associated row in the root relation. In addition, since each relation is part of at most one rooted tree, we hold a single lock per write operation.

\textbf{Physical Locking--} We implement our locking mechanism through lock tables stored in HBase. We create one lock table per root relation. The lock table key has same set of attributes as the root relation's key and it includes a single boolean column that identifies if lock is in use or not. A lock table entry is created when a tuple is inserted into the root table. 

\textbf{Discussion--} We implement light weight hierarchical locking mechanism in Synergy by holding a single lock per write operation. As a downside of hierarchical locking, all rows associated with the root key along all the paths are locked which can affect throughput with concurrent requests trying to grab the lock on the same root key. Note that lock management is not the primary contribution of our work. Other transaction management systems like Themis \cite{Themis}, Tephra \cite{tephra}, Omid \cite{Omid} etc. could also be used.
\subsection{Write Transaction Procedures}
Synergy utilizes transaction procedures to atomically update the base table, views and corresponding indexes upon a base table update. For \textbf{insert} and \textbf{delete} statements, the transaction procedure first acquires the lock on the root key. Then, the base table, applicable views and corresponding indexes are updated using the tuple/key construction procedures described in Section \ref{sec:maintenance}. Finally, the lock is released. Note that each transaction inserts/deletes a single row in/from the base table, applicable views and corresponding indexes.

A base table update may require multi-row updates on the materialized view. Now, while a view is being updated upon a base table update, conflict with the concurrent writes is prevented by the locking mechanism; however, a concurrent read may read dirty data. Hence, to facilitate the detection of a dirty read, we mark the data in views and view-indexes before update and un-mark after update. If dirty data is read in a transaction, then the read is restarted. The \textbf{update} transaction is a 6-step procedure: 1) We first acquire a lock on the root key. 2) Then, we read all the rows that need to be updated. 3) Next, we mark all the rows that need to be updated. 4) Then, we issue a sequence of updates. 5) Next, we un-mark all the updated rows. 6) Finally, we release the lock.

The plan generator component (see Figure \ref{fig:arch}) in the Synergy transaction layer auto generates the execution plan for each write transaction.
\subsection{Transaction Isolation Level}
The Synergy system is restricted to single statement transactions. In addition, Synergy does not support queries in which a relation is used more than once due to potential dirty reads. The Synergy system provides \textbf{ACID semantics} with \textbf{read committed} transaction isolation level, which is also the default transaction isolation level for PostgreSQL \cite{postgres-isolation}. 

A single row is inserted/deleted into/from the base table, applicable views and corresponding indexes upon a base table \textbf{insert/delete}. In addition, to answer a query either a table is used directly or a view involving the table is used but not both. Hence, a reader either reads the entire row or the row is absent from the read result set. This enables read committed behavior for insert and delete statements. 

Recall that the system marks the rows to be updated in a view as dirty before issuing updates, and if a concurrent \textit{scan} reads a dirty row, the scan is restarted. Hence we modify the scan behavior to check for marked rows in the scanned result-set and \textbf{re-scan} if a marked row is present. This ensures that \textbf{update} statement preserves the read committed semantics.

Note that the \textbf{read committed} semantics are preserved during a \textbf{failure scenario}, since the base table lock is held until the system recovers from the failure.  

\begin{figure}[t]
	\centering
		\includegraphics[trim = 0mm 160mm 175mm 10mm, clip, width=.4\textwidth]{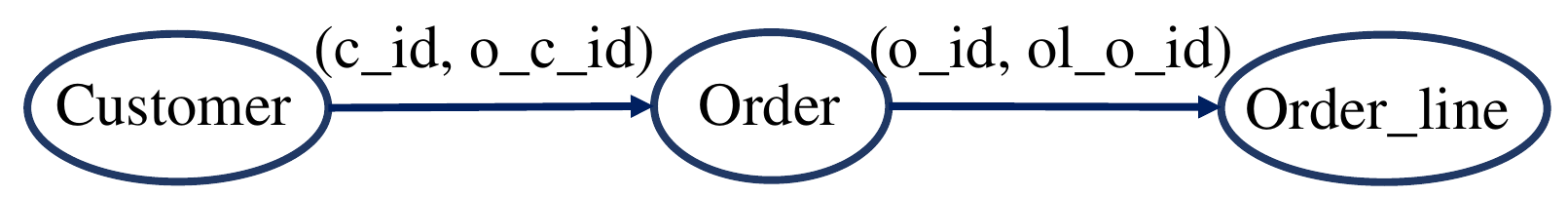}
	\caption{Micro benchmark schema graph.}
	\label{fig:microView}
\end{figure}
\section{Experimental Evaluation}
\label{sec:exp}
In this section we first describe our experiment environment. Next, we use a TPC-W micro-benchmark to evaluate the join performance in HBase. Thereafter, we profile the performance overhead of two phase row locking in HBase. Finally, we evaluate the performance of Synergy system and compare it with four other systems using the full TPC-W benchmark.
\subsection{Experiment Environment}
\subsubsection{Testbed}
\label{sec:testbed}
Amazon EC2 represents our experiment environment. We create an eight node cluster using m4.4xlarge virtual machine (VM) instances. Each instance is configured with 16 vCPU's, 64GB RAM and 120 GB SSD elastic block storage (EBS), running Ubuntu 14.04.

\textit{HBase, HDFS and Zookeeper:} The HDFS NameNode, the HBase HMaster, and the ZooKeeper server processes run on one instance. We designate five instances as slaves, each running the HDFS DataNode and the HBase RegionServer processes. We use Hadoop v2.6.5, HBase v1.2.4.

\textit{Synergy and Phoenix:} We dedicate one instance to host a Synergy transaction layer slave and the Phoenix-Tephra server. Synergy transaction layer master is hosted on the same node that hosts HBase and HDFS masters. We use Phoenix v4.8.2. 

\textit{VoltDB:} We create a five instance VoltDB (v6.8) cluster by hosting a VoltDB daemon on each instance that is also hosting the HDFS DataNode and the HBase RegionServer processes. 

\textit{Client:} We reserve one node as client to drive the workload for each system.
\subsubsection{Performance Metric}
The request response time represents our performance metric, denoted as $\tau$. We measure $\tau$ in the client. 
%
\begin{figure}[t]
	\centering
		\includegraphics[trim = 30mm 2mm 30mm 5mm, clip, width=.5\textwidth]{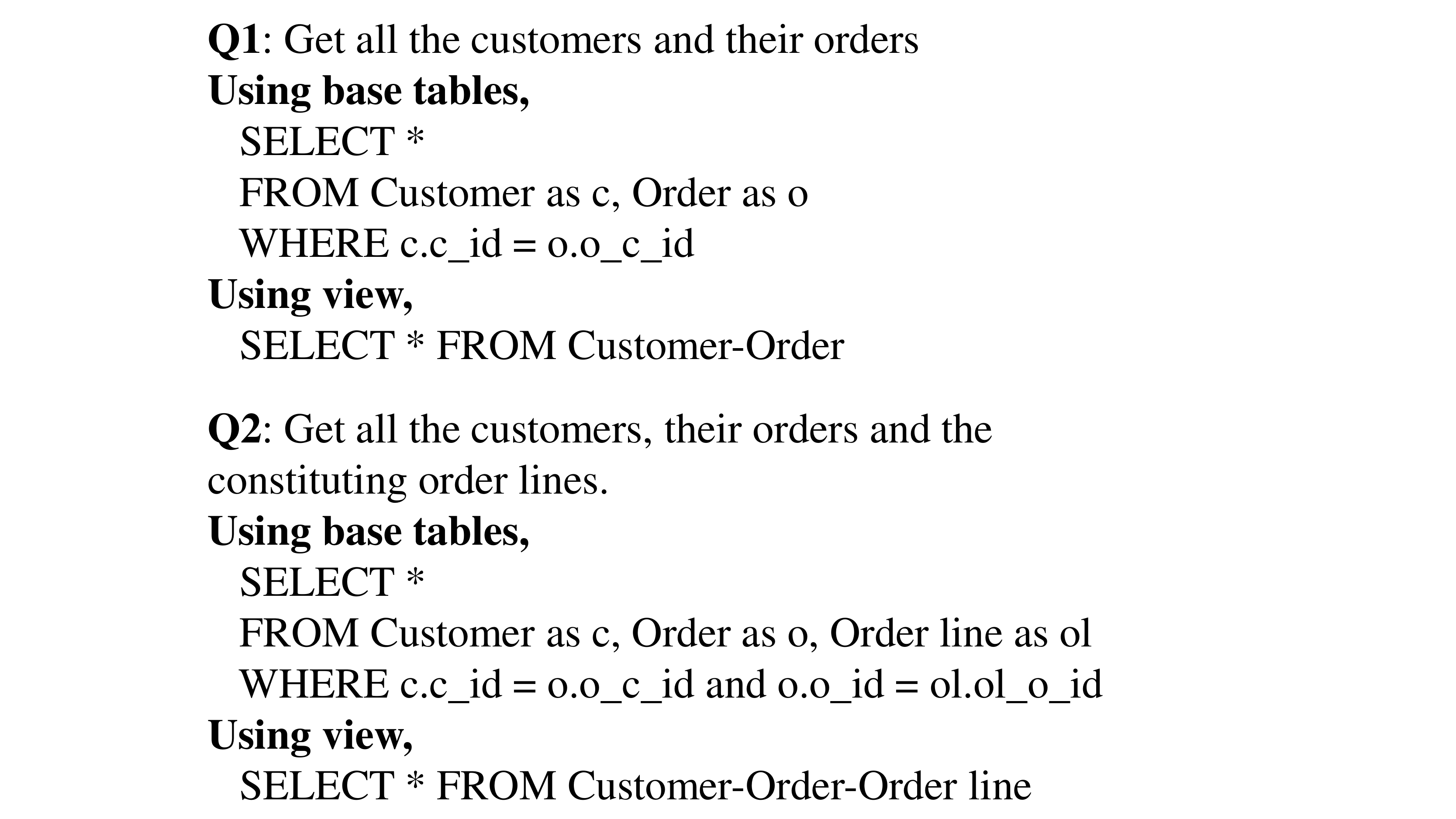}
	\caption{Micro-Benchmark Workload.}
	\label{fig:mb-work}
\end{figure}
\begin{figure}[t]\centering
\hspace*{-.15 in}
\begin{minipage}[t]{0.4\textwidth}\centering
\subfigure[{\small Q1}] {
\includegraphics[width=\linewidth]{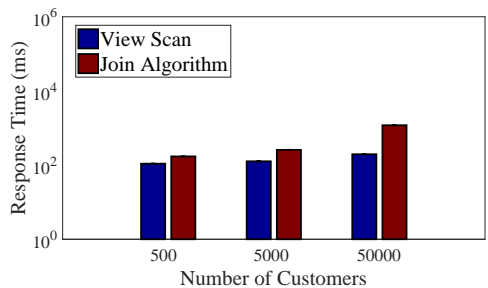}
\label{fig:q1s1}
}
\end{minipage}
\begin{minipage}[t]{0.4\textwidth}\centering
\subfigure[{\small Q2}] {
\includegraphics[width=\linewidth]{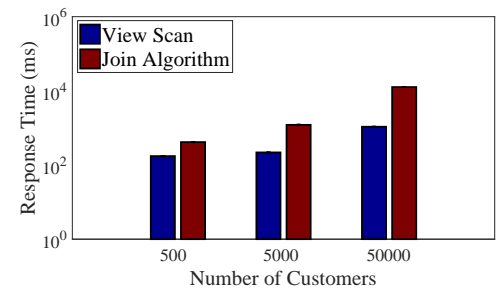}
\label{fig:q2s1}
}
\end{minipage}
\caption{Micro benchmark results to show that performance of join algorithms is slow in HBase. Y axis is drawn at log scale.}
\label{fig:micro}
\end{figure}
%
\begin{figure}[t]
	\centering
		\includegraphics[trim = 12mm 148mm 140mm 12mm, clip, width=.45\textwidth]{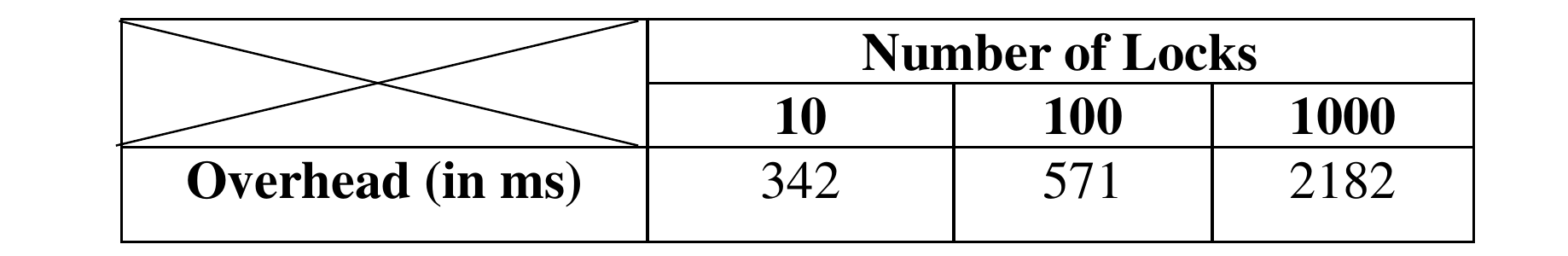}
	\caption{Experiment to show overhead associated with two phase row locking in HBase.}
	\label{fig:lock-overhead}
\end{figure}
\subsection{Micro Benchmark Evaluation}
\label{sec:microb}
We use a TPC-W micro-benchmark to evaluate the join performance in HBase.
\subsubsection{Schema and Workload} 
\label{sec:s+w}
The micro benchmark schema comprises of three relations from the TPC-W benchmark: Customer, Order and Order_line. Customers can have one or more orders and each order can have one or more order lines. Figure \ref{fig:microView} depicts the schema graph for the benchmark schema. Next, to evaluate the join performance, we create a synthetic workload comprising of two foreign key equi-join queries: Q1 (\textit{Customer,Order}) and Q2 (\textit{Customer,Order,Order_line}). 

A join query can be evaluated using two different approaches: 1) using a join algorithm that combines the matching tuples from the specified tables and 2) scanning pre-computed and stored results from a materialized view. Hence, to compare the join algorithm performance with the view scan performance, we materialize the joins in the workload as views. Customer-Order and Customer-Order-Order\_line represent the MVs corresponding to the join queries Q1 and Q2 respectively. Figure \ref{fig:mb-work} presents the workload queries written using base tables and MVs. 
\subsubsection{Experiment Setup and Results}
Each experiment is characterized by the database scale and the join query. We set the cardinality ratio between relations as 1:10. We scale the database by increasing the number of customers in multiples of 10, starting at 500. For each database scale, we major compact both base tables and views after database population. Section \ref{sec:s+w} presents the join queries in the workload. We repeat each experiment 10 times and report the mean and the standard error of response time. Figure \ref{fig:micro} depicts the experiment results with Y axis drawn at log scale.

For the database populated with 50K customers, view scan is 6x and 11.7x faster than the join algorithm for queries Q1 and Q2 respectively. \textit{\textbf{In conclusion, micro-benchmark results show that the join algorithm performance is slow in HBase, providing the motivation for join materialization.}}

\subsection{Locking Overhead Evaluation}
\label{sec:lock-overhead}
In this experiment, our goal is to evaluate the performance overhead of acquiring and releasing row locks in HBase. We create a single lock table in the HBase layer with two attributes: \textit{id} and \textit{lock_status}. The \textit{lock_status} is a boolean column that identifies whether lock is in use or not. We use \textit{checkAndPut} HBase operation in the client node to acquire and release locks. We increase the number of locks in multiples of 10 starting at 10 and measure the overhead in client. We repeat each experiment 10 times and present the mean overhead time. Figure \ref{fig:lock-overhead} shows the experiment results.

Locking overhead with 100 locks is 1.3x the response time of statement W13 in the Synergy system (see Section \ref{sec:write-overhead} and Figure \ref{fig:writes}); W13 represents the most expensive write transaction in the Synergy system. Also note that 100 represents a modest number of locks for a write transaction, since multiple tables with varying cardinalities may be joined together as a view.\textit{\textbf{ In conclusion, overhead associated with the acquisition and release of row locks represents a major transaction performance bottleneck in HBase, motivating the use of a single lock per transaction.}} 
\begin{figure*}
	\centering
		\includegraphics[width=.8\textwidth]{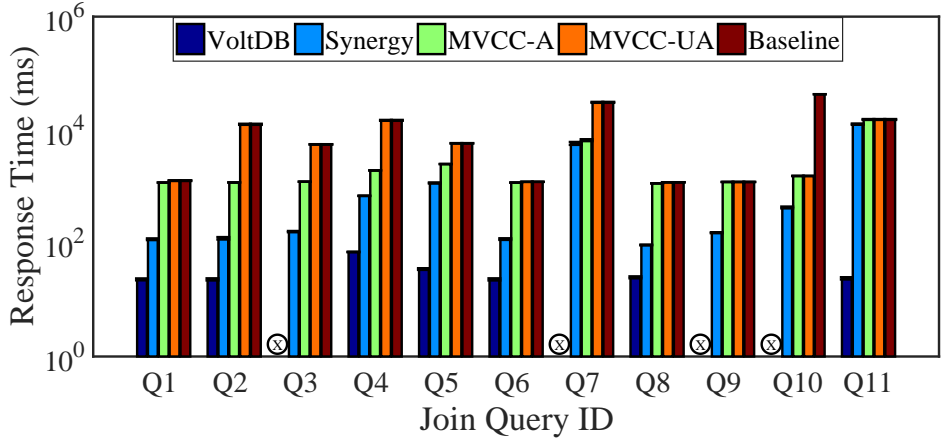}
	\caption{Evaluation and comparison of join performance across different systems using join queries in the TPC-W benchmark. Y axis is drawn at log scale. Join queries \{Q$_3$, Q$_7$, Q$_9$, Q$_{10}$\} are not supported in VoltDB.}
	\label{fig:joins}
\end{figure*}
\subsection{TPC-W Benchmark Evaluation}
\subsubsection{Benchmark}
\label{sec:bench}
TPC-W \cite{tpcw} is a transactional web benchmark. It has a two tier architecture including a web tier and a database tier. TPC-W workload includes 14 different types of web requests where each request is modeled as a servlet. Each servlet is in turn composed of one or more SQL statements. We analyze the TPC-W servlets to extract all the SQL statements that can be invoked at the database tier. Extracted set of SQL statements represents our workload.

We exclude a DELETE statement (DELETE FROM \textit{shopping_cart_line} WHERE \textit{scl_sc_id} = ?) from the workload that may affect multiple base table rows. Phoenix currently does not provide an implementation of the \textit{soundex} algorithm; hence, we exclude two join queries from the workload that use \textit{soundex} algorithm.

The database size (DB$_{size}$) can be modulated by varying two parameters: the number of customers (NUM\_CUST) and the number of items (NUM\_ITEMS).  We set NUM\_ITEMS to 10 * NUM\_CUST. In addition, we change the cardinality between the Customer and the Orders table from .9 to 10. We populate the database with 1 million customers. For each system that utilizes HBase as the storage layer, we major compact base tables, indexes and MVs after the database population. 

\begin{figure}[t]
	\centering
		\includegraphics[trim = 12mm 108mm 55mm 5mm, clip, width=.48\textwidth]{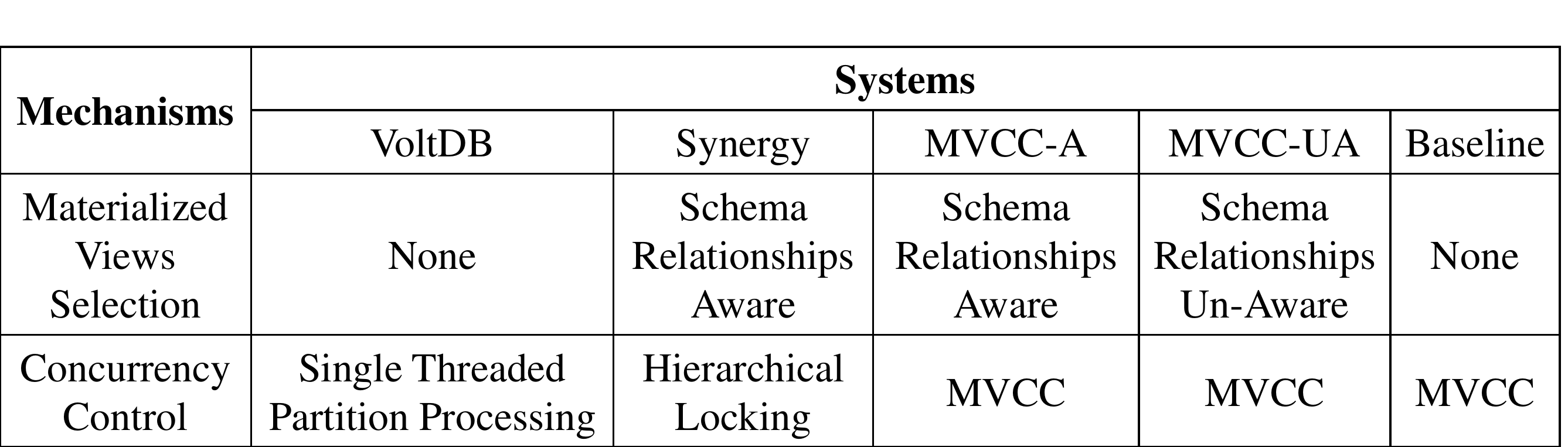}
	\caption{Materialized views selection mechanism and concurrency control mechanism used in each evaluated system.}
	\label{fig:systems}
\end{figure}

\subsubsection{Systems Evaluated}
\textit{Synergy:} We use Q$_{TPC-W}$ = \{\textit{Author, Customer, Country}\} as the roots set to generate views in the Synergy system. We create base tables, selected views and corresponding indexes in HBase. In addition, we create lock tables for each root in Q$_{TPC-W}$ . We disable the Phoenix-Tephra transaction support.  

\textit{MVCC-UA:} To compare our views generation and selection mechanism with \cite{Agrawal}, we deploy SQL Server 2012 on a single EC2 VM instance. Next, we populate the TPC-W database with 1 million customers and run the TPC-W benchmark queries. Then, we use the SQL Server's database engine tuning advisor to analyze the profiled workload and generate views. We create the generated views along with base tables and indexes in HBase and run the workload with Phoenix-Tephra transaction support (MVCC) enabled.

\textit{MVCC-A:} In addition to the base tables and indexes, we create the views and the view-indexes generated by the Synergy system in HBase and run the workload with Phoenix-Tephra transaction support (MVCC) instead of the specialized transaction support used in Synergy.

\textit{Baseline:} We only create base tables and corresponding indexes in HBase and run the workload with Phoenix-Tephra transaction support (MVCC).

\textit{VoltDB:} A VoltDB table can either be partitioned or replicated. The partitioning column is specified by the user and partitioned tables can only be joined on equality of partitioning column. Now, a table can join with other tables using different columns in different queries of the workload; however, since each table can only be partitioned on a single column, only a subset of workload join queries may work for a partitioning scheme. 

To profile the performance of maximum number of joins in the TPC-W benchmark we use \textbf{three different partitioning schemes in VoltDB}. However, note that in practice only one partitioning scheme could be used for a database. Also, note that only base tables and corresponding indexes are used in VoltDB.

Figure \ref{fig:systems} summarizes the MVs creation and concurrency control mechanisms used in each evaluated system.
\subsubsection{Performance Evaluation of Joins in the TPC-W Benchmark}
\label{subsec:joins}
\textit{Experiment setup--}
In this set of experiments, we evaluate and compare the join performance across different systems using the join queries in the TPC-W benchmark. Recall that we used three different partitioning schemes in VoltDB to support maximum number of TPC-W joins, using any single partitioning scheme less than 50\% of the TPC-W joins are supported.

We evaluate each query10 times and present the mean and the standard error of the recorded response times. See appendix for the specification of join queries in the TPC-W benchmark. Figure \ref{fig:joins} presents the experiment results. Note that join queries \{Q$_3$, Q$_7$, Q$_9$, Q$_{10}$\} are not supported in VoltDB.

\textit{Discussion--} On an average the join queries in Synergy are 19.5x, 6.2x and 28.2x faster as compared to the MVCC-UA, MVCC-A and Baseline system respectively. The view selection mechanism in the Synergy system selects more MVs as compared to MVCC-UA, resulting in significantly larger join performance benefit. In MVCC-UA, the response time of Q$_{10}$ is significantly lower than the Baseline system since MVCC-UA utilizes a materialized view for query evaluation. The join performance in Synergy system with specialized concurrency control is marginally better than MVCC-A that uses MVCC. The join queries that used views in Synergy are on an average 11x slower than VoltDB (excluding queries that are not supported in VoltDB). \textit{\textbf{In conclusion, join response times in the Synergy system with selected views are significantly lower as compared to MVCC-UA and Baseline system for the benchmark queries. In addition, although the Synergy join performance is slower than VoltDB, Synergy allows for significantly more expressive joins than VoltDB.}} 
\begin{figure*}
	\centering
		\includegraphics[width=.8\textwidth]{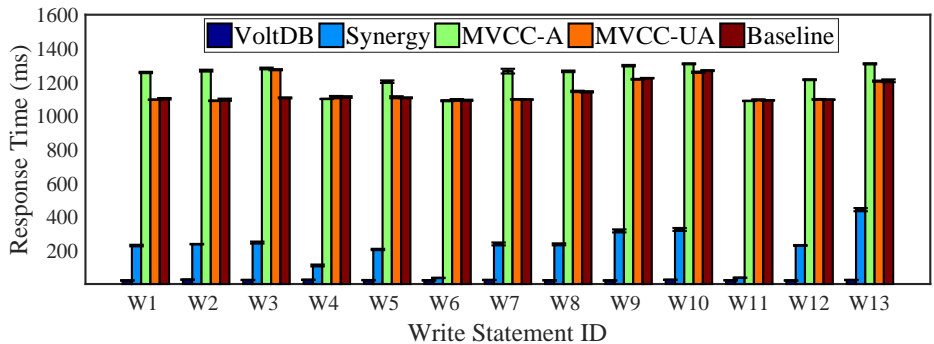}
	\caption{Performance Evaluation of the write statements in the TPC-W benchmark to exhibit the overhead of lock management and updating views in the Synergy system. Comparison of write statement performance across different systems.}
	\label{fig:writes}
\end{figure*}
\subsubsection{Performance Evaluation of Write Statements in the TPC-W Benchmark}
\label{sec:write-overhead}
\textit{Experiment setup--}In this set of experiments, we aim to evaluate the performance overhead of acquiring/releasing a lock and updating MVs in the Synergy system. In addition, we compare the write statement performance across different systems using the write statements in the TPC-W benchmark. See appendix for the specification of write statements in the TPC-W benchmark. We evaluate each statement 10 times and present the mean and the standard error of the recorded response times. Figure \ref{fig:writes} presents our experiment results.

\textit{Discussion--}On an average the write statements in Synergy are 9x, 8.6x and 8.6x less expensive than MVCC-UA, MVCC-A and Baseline system respectively. In Synergy system, the execution time of statements W6 and W11 is significantly lower than the other write statements since the corresponding relation is not part of any views. Although Baseline system does not use any MVs and MVCC-UA utilizes only one materialized view, the statement response times in these systems are high since MVCC adds an overhead of 800-900 ms to each statement's execution time. On an average the write statements in Synergy are 9.4x more expensive than the VoltDB. \textit{\textbf{In conclusion, experimental results show that the use of hierarchical locking in Synergy system significantly reduces the write transaction response times in presence of MVs.}}
\begin{table}[ht]
\centering
\caption{Sum of RT of all the statements in the TPC-W benchmark to quantify trade off between read performance gain and write performance overhead of using MVs in each evaluated system. VoltDB is excluded since it does not support all queries in the benchmark.}
\label{fig:all_stmt}
\scalebox{0.8}{
\begin{tabular}{|l|c|c|c|c|}
\hline
\multirow{2}{*}{}                                                                                & \multicolumn{4}{c|}{\textbf{Evaluated Systems}} \\ \cline{2-5} 
                                                                                                 & \textbf{Synergy}  & \textbf{MVCC-A} & \textbf{MVCC-UA} & \textbf{Baseline} \\ \hline
\multicolumn{1}{|c|}{\begin{tabular}[c]{@{}c@{}}\textbf{Mean Response Time}\\  (in seconds)\end{tabular}} & 33.7     & 77.4   & 132.4   & 173.4    \\ \hline
\multicolumn{1}{|c|}{\textbf{Standard Error}}                                                             & .03      & .02    & .06     & .07      \\ \hline
\end{tabular}}
\end{table}
\begin{table}[]
\centering
\caption{Database sizes across different evaluated systems.}
\label{fig:dbsize}
\scalebox{0.7}{
\begin{tabular}{|c|c|c|c|c|c|}
\hline
\multirow{2}{*}{\textbf{No. of Customers}} & \multicolumn{5}{c|}{\textbf{Database Size (in GB)}}         \\ \cline{2-6} 
                                  & \textbf{VoltDB} & \textbf{Synergy}  & \textbf{MVCC-A} & \textbf{MVCC-UA} & \textbf{Baseline}   \\ \hline
 \textbf{1M}                           & 31.8   &    92     & 91.8     & 45.73   &  43.8  \\ \hline
\end{tabular}}
\end{table}
\subsubsection{Performance Comparison of All Evaluated Systems}
\textit{Experiment setup--}In this set of experiments, we evaluate the performance gain and the storage overhead of using MVs in the Synergy system and compare it with the other systems. Note that we exclude VoltDB since it does not support all join queries in the TPC-W benchmark. We evaluate the performance of systems using all the statements in the TPC-W benchmark.

During an experiment, we run each benchmark SQL statement and record its response time. Next, we compute the sum of response time of all statements. We run each experiment 10 times and present mean and standard error of the benchmark response time. Table \ref{fig:all_stmt} presents the experiment results. Table \ref{fig:dbsize} summarizes the database sizes across different systems.

\textit{Discussion--} Synergy system exhibits a performance improvement of 74.5\%, 56.3\% and 80.5\% as compared to the MVCC-UA, MVCC-A and Baseline system respectively. Conversely, the database size in the Synergy system is 2x, 1x and 2.1x the database size in the MVCC-UA, MVCC-A and Baseline system respectively. Hence, Synergy system trades slight write performance degradation and increased disk utilization for faster join performance. \textit{\textbf{In conclusion, the specialized concurrency control mechanism and the MVs generation mechanism in the Synergy system significantly improve the read performance without shifting the bottleneck to the write performance.}}

\section{Related Work}
\label{sec:rw}
\textbf{Materialized Views.} MVs have been studied from multiple standpoints in the SQL domain: view maintenance, view matching, automated views selection, dynamic view maintenance etc. In \cite{Blakeley86, Quass, Blakeley89, asyncView} authors explore the problem of efficient view maintenance in response to the base table updates. The dynamic views \cite{Zhou} introduce storage efficiency by automatically adapting the number of rows in the view in response to the changing workload. The view matching techniques are utilized in query optimization to determine the query containment and the query derivability \cite{Larson, Yang, Goldstein}. In \cite{Agrawal}, authors propose a workload driven mechanism to automate the task of selecting an optimal number of views and indexes for decision support system applications. The MVs selection and maintenance in a transaction processing NoSQL data store raises novel challenges since most of the existing views selection approaches are oblivious to the relationship between schema relations which can lead to heavy view maintenance costs and can shift the bottleneck from reads to writes. To this end, Synergy proposes a novel, schema relationships aware view selection mechanism.  

\textbf{Data Partitioning.} Megastore \cite{Megastore}, F1 \cite{Shute} and Elastras \cite{Das2} harness hierarchical schema structure to cluster related data together and minimize the distributed transactions. On the contrary, Synergy generates MVs utilizing hierarchical schema structure to reduce query run times. In \cite{Bin}, authors automate the task of data partitioning by developing a technique for automated selection of root relations in a schema. Schism \cite{Curino} proposes fine grained data partitioning by co-locating related tuples based on workload logs.
 
\textbf{Transactions.} Transaction support in the majority of the first generation NoSQL stores \cite{HBase, Accumulo} and Big Data systems \cite{asterix} is limited to single-keys. G-Store \cite{Das1} extends HBase to support multi-key transactions in a layer on top using a 2 phase locking protocol. Similar to G-Store, we implement write transactions in a layer on top of HBase. CloudTPS \cite{Cloudtps} supports multi-key read/write transactions in a highly scalable DHT based transaction layer using optimistic concurrency control (OCC). In \cite{Wei}, authors extend CloudTPS to support consistent foreign-key equi-joins. ecStore \cite{Vo} provides snapshot isolation using MVCC based on the global timestamp transaction ordering in a decoupled layer on top of an ordered key-value store BATON. ElasTras \cite{Das2} proposes a novel key-value store that implements MVCC based transactions. Percolator \cite{Peng} extends Bigtable to allow cross-row, cross-table ACID transactions and enables incremental updates to the web index data stored in BigTable. Megastore \cite{Megastore} introduces entity groups as a granule of physical data partitioning and supports ACID transactions with in an entity group. F1 \cite{Shute} is built on top of Spanner \cite{Spanner} and supports global ACID transactions for the Google AdWords business. In contrast with Spanner, Synergy is limited to single data center use; however, Synergy enables enhanced SQL query expressiveness and does not require sophisticated infrastructure including atomic clocks, GPS etc. The NewSQL databases \cite{hstore, voltdb} scale out linearly while ensuring ACID semantics; however, the join support is limited to partitioning keys. The first generation of NewSQL systems required all data to reside in main memory; however, recent work \cite{anticache} overcomes this limitation by keeping cold data on the disk.  

\section{Conclusions}
\label{sec:conclusion}
In this paper we present the Synergy system, a data store that leverages schema based--workload driven materialized views and a specialized concurrency control system on top of a NoSQL database that allows for scalable data management with familiar relational conventions. Synergy trades slight write performance degradation and increased disk utilization for faster join performance (compared to standard NoSQL databases) and improved query expressiveness (compared to NewSQL databases). Experiment results on a lab cluster using the TPC-W benchmark show the efficacy of our system. 

\section{Acknowledgement}
\label{sec:ack}
This work was supported in part by National Science Foundation awards 1526014 and 1150169.

{\scriptsize 
{
\bibliographystyle{IEEEtran}
\bibliography{IEEEabrv,transformer}

\begin{thebibliography}{10}
\providecommand{\url}[1]{#1}
\csname url@samestyle\endcsname
\providecommand{\newblock}{\relax}
\providecommand{\bibinfo}[2]{#2}
\providecommand{\BIBentrySTDinterwordspacing}{\spaceskip=0pt\relax}
\providecommand{\BIBentryALTinterwordstretchfactor}{4}
\providecommand{\BIBentryALTinterwordspacing}{\spaceskip=\fontdimen2\font plus
\BIBentryALTinterwordstretchfactor\fontdimen3\font minus
  \fontdimen4\font\relax}
\providecommand{\BIBforeignlanguage}[2]{{%
\expandafter\ifx\csname l@#1\endcsname\relax
\typeout{** WARNING: IEEEtran.bst: No hyphenation pattern has been}%
\typeout{** loaded for the language `#1'. Using the pattern for}%
\typeout{** the default language instead.}%
\else
\language=\csname l@#1\endcsname
\fi
#2}}
\providecommand{\BIBdecl}{\relax}
\BIBdecl

\bibitem{kossmann}
D.~Kossmann~et al., ``An evaluation of alternative architectures for
  transaction processing in the cloud,'' in \emph{SIGMOD}, 2010, pp. 579--590.

\bibitem{ycsb}
B.~F. Cooper~et al., ``Benchmarking cloud serving systems with ycsb,'' in
  \emph{SoCC}, 2010, pp. 143--154.

\bibitem{hstore}
R.~Kallman~et al., ``H-store: A high-performance, distributed main memory
  transaction processing system,'' \emph{Proc. VLDB Endow.}, pp. 1496--1499,
  2008.

\bibitem{Fbmsg}
``Storage infrastructure behind {F}acebook messages: Using {HB}ase at scale.
  [online]. available: http://sites.computer.org/debull/a12june/p4.pdf.''

\bibitem{Netflix}
``No{SQL} at {N}etflix. [online]. available:
  http://techblog.netflix.com/2011/01/nosql-at-netflix.html.''

\bibitem{HBase}
``H{B}ase. [online]. available: http://hbase.apache.org/.''

\bibitem{voltdb}
``Volt{DB}. [online]. available: https://voltdb.com/.''

\bibitem{asyncView}
P.~Agrawal~et al., ``Asynchronous view maintenance for vlsd databases,'' in
  \emph{SIGMOD}, 2009, pp. 179--192.

\bibitem{Larson}
P.-A. Larson and H.~Z. Yang, ``Computing queries from derived relations,'' in
  \emph{VLDB}, 1985, pp. 259--269.

\bibitem{Yang}
H.~Z. Yang and P.-A. Larson, ``Query transformation for psj-queries,'' in
  \emph{VLDB}, 1987, pp. 245--254.

\bibitem{Bigtable}
F.~Chang~et al., ``Bigtable: A distributed storage system for structured
  data,'' in \emph{OSDI}, 2006, pp. 15--15.

\bibitem{Accumulo}
``Accumulo. [online]. available: https://accumulo.apache.org/.''

\bibitem{Phoenix}
``Apache {P}hoenix. [online]. available: http://phoenix.apache.org/.''

\bibitem{tephra}
``Thepra. [online]. available: http://tephra.incubator.apache.org/.''

\bibitem{Goldstein}
J.~Goldstein and P.-A. Larson, ``Optimizing queries using materialized views: A
  practical, scalable solution,'' in \emph{SIGMOD}, 2001, pp. 331--342.

\bibitem{Agrawal}
S.~Agrawal~et al., ``Automated selection of materialized views and indexes in
  sql databases,'' in \emph{VLDB}, 2000, pp. 496--505.

\bibitem{cap}
S.~Gilbert and N.~Lynch, ``Brewer's conjecture and the feasibility of
  consistent, available, partition-tolerant web services,'' \emph{SIGACT News},
  2002.

\bibitem{Themis}
``Themis. [online]. available: https://github.com/xiaomi/themis.''

\bibitem{Omid}
``Omid. [online]. available: https://github.com/yahoo/omid.''

\bibitem{postgres-isolation}
``Postgre{SQL}: {T}ransaction {I}solation. [online]. available:
  http://www.postgresql.org/docs/9.1/static/transaction-iso.html.''

\bibitem{tpcw}
``\uppercase{TPC-W} benchmark. [online]. available: http://www.tpc.org/tpcw/.''

\bibitem{Blakeley86}
J.~A. Blakeley, P.-A. Larson, and F.~W. Tompa, ``Efficiently updating
  materialized views,'' in \emph{SIGMOD}, 1986, pp. 61--71.

\bibitem{Quass}
D.~Quass, A.~Gupta, I.~S. Mumick, and J.~Widom, ``Making views
  self-maintainable for data warehousing,'' in \emph{DIS}, 1996, pp. 158--169.

\bibitem{Blakeley89}
J.~A. Blakeley, N.~Coburn, and P.-V. Larson, ``Updating derived relations:
  Detecting irrelevant and autonomously computable updates,'' \emph{ACM Trans.
  Database Syst.}, vol.~14, no.~3, Sep. 1989.

\bibitem{Zhou}
J.~Zhou, P.-A. Larson, J.~Goldstein, and L.~Ding, ``Dynamic materialized
  views,'' in \emph{ICDE}, 2007, pp. 526--535.

\bibitem{Megastore}
J.~Baker~et al., ``Megastore: Providing scalable, highly available storage for
  interactive services,'' in \emph{CIDR}, 2011, pp. 223--234.

\bibitem{Shute}
J.~Shute~et al., ``F1: A distributed sql database that scales,'' \emph{Proc.
  VLDB Endow.}, pp. 1068--1079, 2013.

\bibitem{Das2}
S.~Das~et al., ``Elastras: An elastic, scalable, and self-managing
  transactional database for the cloud,'' \emph{ACM Trans. Database Syst.},
  vol.~38, no.~1, pp. 5:1--5:45, Apr. 2013.

\bibitem{Bin}
B.~Liu~et al., ``Automatic entity-grouping for {OLTP} workloads,'' in
  \emph{ICDE}, 2014.

\bibitem{Curino}
C.~Curino~et al., ``Schism: A workload-driven approach to database replication
  and partitioning,'' \emph{Proc. VLDB Endow.}, pp. 48--57, 2010.

\bibitem{asterix}
S.~Alsubaiee~et al., ``Asterixdb: A scalable, open source bdms,'' \emph{Proc.
  VLDB Endow.}, pp. 1905--1916, 2014.

\bibitem{Das1}
S.~Das~et al., ``G-store: A scalable data store for transactional multi key
  access in the cloud,'' in \emph{SoCC}, 2010, pp. 163--174.

\bibitem{Cloudtps}
Z.~Wei, G.~Pierre, and C.-H. Chi, ``{CloudTPS}: Scalable transactions for {W}eb
  applications in the cloud,'' \emph{IEEE Transactions on Services Computing},
  vol.~5, no.~4, pp. 525--539, Oct-Dec 2012.

\bibitem{Wei}
------, ``Scalable join queries in cloud data stores,'' in \emph{CCGrid}, 2012,
  pp. 547--555.

\bibitem{Vo}
H.~T. Vo~et al., ``Towards elastic transactional cloud storage with range query
  support,'' \emph{Proc. VLDB Endow.}, pp. 506--514, 2010.

\bibitem{Peng}
D.~Peng and F.~Dabek, ``Large-scale incremental processing using distributed
  transactions and notifications,'' in \emph{OSDI}, 2010, pp. 1--15.

\bibitem{Spanner}
J.~C. Corbett~et al., ``Spanner: Google's globally-distributed database,'' in
  \emph{OSDI}, 2012, pp. 251--264.

\bibitem{anticache}
J.~DeBrabant~et al., ``Anti-caching: A new approach to database management
  system architecture,'' \emph{Proc. VLDB Endow.}, pp. 1942--1953, 2013.

\end{thebibliography}
}
\section{Appendix}
\begin{figure}[h]
	\centering
		\includegraphics[trim = 0mm 48mm 45mm 1mm, clip, width=.49\textwidth]{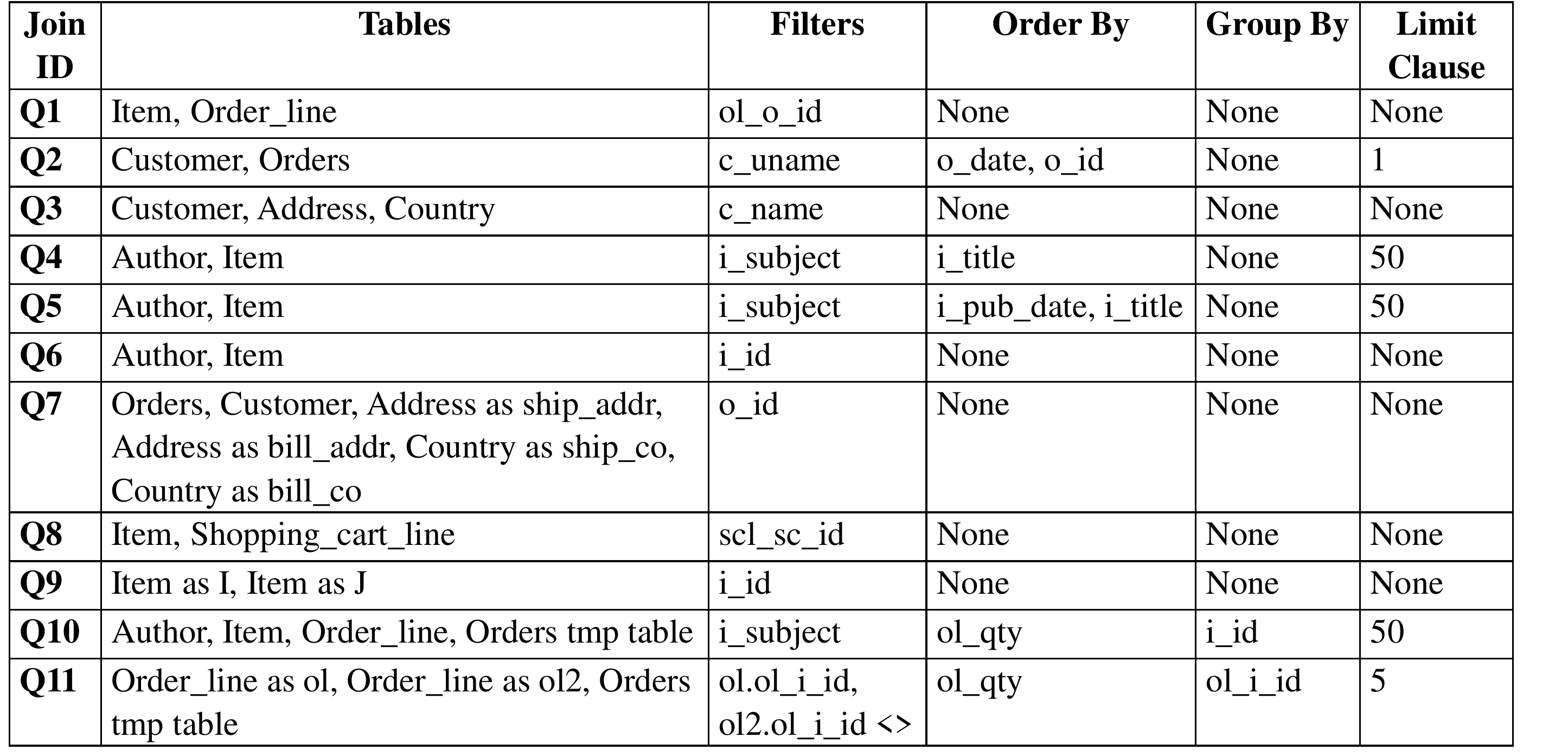}
	\caption{Specification of joins in the TPC-W Benchmark.}
	\label{fig:query-tpcw}
\end{figure}

\begin{figure}[h]
	\centering
		\includegraphics[trim = 1mm 140mm 85mm 1mm, clip, width=.48\textwidth]{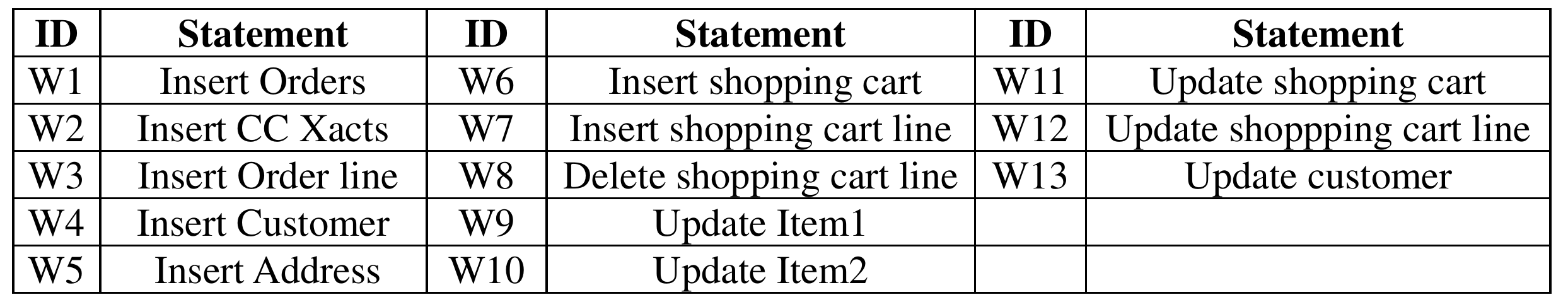}
	\caption{Specification of write statements in TPC-W Benchmark.}
	\label{fig:write-tpcw}
\end{figure}
\end{document}